# Discussion on a Class of Model-Free Adaptive Control for Multivariable Systems

Feilong Zhang

***Abstract*—To fully understand and effectively implement model-free adaptive control (MFAC), we develop a class of MFAC for multivariable systems and analyze its stability using the closed-loop function.**

## I. INTRODUCTION

A significant amount of work concerning MFAC has been published in this decade. [6] has discussed some problems in [1]-[5] that designed MFAC in single-input single-output systems. Similar to [6], we have analyzed several problems in multiple-input multiple-output (MIMO) systems as follows.

i) The estimated leading coefficient matrix of the control input vector is forcibly restricted to a special kind of diagonally dominant matrix with the unchangeable sign of the diagonal elements [3]-[5]. This paper and [7] argue that it is crucial for the estimated leading coefficient matrix to accurately reflect the leading coefficients of the actual model. If the signs of the estimated leading coefficients are opposite to those of the actual model, it can easily result in divergence. In addition, the pseudo orders of the current equivalent dynamic linearization model and MFAC are limited to $1\leq L_y\leq n_y$ and $1\leq L_u\leq n_u$. However, it seems that the most important case is $L_y=n_y+1$ and $L_u=n_u+1$ with the purpose of the pseudo-Jacobin matrix acting as the real Jacobin matrix. To this end, we extend the range to $0\leq L_y$ and $1\leq L_u$.

ii) The static error of the unit-ramp response of the system is eliminated by choosing $\lambda=0$ when $M_u\geq M_y$, which is shown in this brief. This finding contradicts the results presented in [1]-[5], which demonstrated that the tracking error of the system controlled by MFAC converges to zero under the condition that $\lambda$ is large enough. Moreover, [6] explains why the tracking error of the step response of the system controlled by the current MFAC converges to zero, which is not determined by $\lambda$ and should be attributed to the inherent presence of an integrator.

iii) The current works on the MFAC method in MIMO systems take the norm of the inverse matrix in the controller design just for analyzing the system stability through the contraction mapping method. However, this will lead to inaccurate input and output coupling information in the controller.

iv) The notion of model-free presented for nonlinear systems in [1]-[5] is not straightforward for operating engineers to understand correctly and to master its essence. [7] analyzes this class of controller in linear deterministic finite-dimensional systems in simulations to illustrate its working principle and discuss the nature of this kind of adaptive controller.

Manuscript received Dec 3, 2020. This work was supported in part by the xxxxxxxxxx.

Feilong Zhang is with the State Key Laboratory of Robotics, Shenyang Institute of Automation, Chinese Academy of Sciences, Shenyang 110016, China (e-mail: zhangfeiong@sia.cn).

On the other hand, some notable advantages of the proposed method are as follows.

[9]-[14] decompose autoregressive and moving average model (NARMAM) or industrial process nonlinear model into a simple linear model and an unmolded dynamics (UD) around an operating point. Then, the corresponding controller with compensation of UD is designed and analyzed using the closed-loop system equation. Compared to these works, the equivalent-dynamic-linearization-model (EDLM) offers the advantage of describing NARMAM at any points, making it easier and more widely applicable for the MFAC. Additionally, can anyone achieve the control performance of Example 1 within 15 minutes? After mastering the proposed method, one can easily achieve that.

Some extraordinary contributions are shown as follows.

The full-form equivalent-dynamic-linearization model is extended to a more general form by incorporating the pseudo-Jacobian matrix (PJM), which is no longer restricted by the assumption of a special diagonally dominant matrix. Additionally, this extension allows for a difference in the number of input variables and output variables in the system. As a result, the MFAC can be applied more extensively. This is all possible because we have analyzed the system performance using the closed-loop system equations and the static error analysis rather than relying on the current contraction mapping method. Along with this, we have concluded that the static error of the system following the reference $k^n\boldsymbol{I}(n\geq 1)$ is positively correlated with $\lambda$ in this brief, [6] and [8]. These imply that the conclusion about the convergence of tracking errors in current works may not be reasonable.

Only when we study the MFAC in a reasonable way can we apply this method correctly. This is one of the main motivations in this brief. Finally, we have used iterative MFAC to design a simple yet useful robotic controller that demonstrates excellent performance.

## II. EQUIVALENT DYNAMIC LINEARIZATION MODEL FOR MULTIVARIABLE SYSTEMS

This section presents a full-form equivalent dynamic linearization model for a family of multivariable systems, which is used for MFAC controller design and analysis in the next section.

The MIMO nonlinear system is given as:

$$\boldsymbol{y}(k+1)=\boldsymbol{f}(\boldsymbol{y}(k),\cdots,\boldsymbol{y}(k-n_y),\boldsymbol{u}(k),\cdots,\boldsymbol{u}(k-n_u)) \tag{1}$$

where $\boldsymbol{f}(\cdots)=\left[f_1(\cdots),\cdots,f_{My}(\cdots)\right]^T\in\boldsymbol{R}^{My}$ is assumed to be the

nonlinear vector-valued function; $\boldsymbol{u}(k)$ and $\boldsymbol{y}(k)$ are the system's input vector and output vector, respectively; $n_u+1$, $n_y+1 \in \mathrm{Z}$ are the corresponding orders. The dimensions of $\boldsymbol{y}(k)$ and $\boldsymbol{u}(k)$ are $M_y$ and $M_u$, respectively.

Define

$$\boldsymbol{\varphi}(k)=[\boldsymbol{y}(k),\cdots,\boldsymbol{y}(k-n_y),\boldsymbol{u}(k),\cdots,\boldsymbol{u}(k-n_u)] \tag{2}$$

to rewrite (1) as

$$\boldsymbol{y}(k+1)=\boldsymbol{f}(\boldsymbol{\varphi}(k)) \tag{3}$$

*Assumption* 1*:* The partial derivatives of $\boldsymbol{f}(\cdots)$ concerning all input and output vectors are continuous.

*Theorem 1:* Given system (1) satisfying *Assumptions* 1, if $\Delta\boldsymbol{H}(k)\neq\boldsymbol{0}$, $0\leq L_y$ and $1\leq L_u$, there must exist a matrix $\boldsymbol{\phi}_L^T(k)$ named pseudo Jacobian matrix, and (1) can be transformed into the following model:

$$\Delta\boldsymbol{y}(k+1)=\boldsymbol{\phi}_L^T(k)\Delta\boldsymbol{H}(k) \tag{4}$$

where

$$\begin{aligned}\boldsymbol{\phi}_L^T(k)&=[\boldsymbol{\phi}_{Ly}^T(k)_{My\times(Ly\bullet My)},\boldsymbol{\phi}_{Lu}^T(k)_{My\times(Lu\bullet Mu)}]\\&=[\boldsymbol{\Phi}_1(k),\cdots,\boldsymbol{\Phi}_{Ly}(k),\boldsymbol{\Phi}_{Ly+1}(k),\cdots,\boldsymbol{\Phi}_{Ly+Lu}(k)]\end{aligned},$$

$$\boldsymbol{\Phi}_i(k)=\begin{bmatrix}\phi_{11i}(k) & \phi_{12i}(k) & \cdots & \phi_{1Myi}(k)\\ \phi_{21i}(k) & \phi_{22i}(k) & \cdots & \phi_{2Myi}(k)\\ \vdots & \vdots & \vdots & \vdots\\ \phi_{My1i}(k) & \phi_{My2i}(k) & \cdots & \phi_{MyMyi}(k)\end{bmatrix}\in\boldsymbol{R}^{My\times My},\ i=1,\cdots,L_y;$$

Similarly, $\boldsymbol{\Phi}_i(k)\in\boldsymbol{R}^{My\times Mu}$, $i=L_y+1,\cdots,L_y+L_u$;

$\boldsymbol{H}(k)=[\boldsymbol{y}^T(k),\cdots,\boldsymbol{y}^T(k-L_y+1),\boldsymbol{u}^T(k),\cdots,\boldsymbol{u}^T(k-L_u+1)]^T$ is a vector containing the system control input and output vectors within the time window $[k-L_u+1,k]$ and $[k-L_y+1,k]$, respectively. The integers $0\leq L_y$ and $1\leq L_u$ are named pseudo orders of the system and $\Delta\boldsymbol{H}(k)=\boldsymbol{H}(k)-\boldsymbol{H}(k-1)$.

*Remark 1*: [3] gives the proof of *Theorem 1* in the case of $1\leq L_y\leq n_y$, $1\leq L_u\leq n_u$ and we further prove *Theorem 1* for the orders $0\leq L_y$ and $1\leq L_u$ in Appendix.

We prefer $L_y=n_y+1$ and $L_u=n_u+1$ in applications when $n_y$ and $n_u$ are obtainable. If not, we typically select appropriate $L_y$ and $L_u$ values that satisfy $n_y+1\leq L_y$ and $n_u+1\leq L_u$ in adaptive control. One rationale behind this is that the online estimated coefficient matrices of redundant terms $\Delta\boldsymbol{y}(k-n_y-1),\cdots,\Delta\boldsymbol{y}(k-L_y+1)$ and $\Delta\boldsymbol{u}(k-n_u-1),\cdots,\Delta\boldsymbol{u}(k-L_u+1)$ may approach zero matrix values. Concurrently, the estimated coefficients of $\Delta\boldsymbol{y}(k),\cdots,\Delta\boldsymbol{y}(k-n_y)$ and $\Delta\boldsymbol{u}(k),\cdots,\Delta\boldsymbol{u}(k-n_u)$ tend to be closer to their actual values compared to when $0\leq L_y\leq n_y$ and $1\leq L_u\leq n_u$.

## III. FFDL-MFAC DESIGN AND STABILITY ANALYSIS

This section provides the design and stability analysis methods for MFAC.

### *A. Design of Model-Free Adaptive Control*

We rewrite (4) into (5).

$$\boldsymbol{y}(k+1)=\boldsymbol{y}(k)+\boldsymbol{\phi}_L^T(k)\Delta\boldsymbol{H}(k) \tag{5}$$

A control input criterion function is given as:

$$J=\left[\boldsymbol{y}^*(k+1)-\boldsymbol{y}(k+1)\right]^T\left[\boldsymbol{y}^*(k+1)-\boldsymbol{y}(k+1)\right]+\Delta\boldsymbol{u}^T(k)\boldsymbol{\lambda}\Delta\boldsymbol{u}(k) \tag{6}$$

where $\boldsymbol{\lambda}=dig(\lambda_1,\cdots,\lambda_{Mu})$ is the weighted diagonal matrix and $\lambda_i$ ($i=1,\cdots,M_u$) are equal for the system analysis by [3]; $\boldsymbol{y}^*(k+1)=\left[y_1^*(k+1),\cdots,y_{My}^*(k+1)\right]^T$ is the desired trajectory vector.

We substitute (5) into (6) and solve the optimization condition $\partial J/\partial\Delta\boldsymbol{u}(k)=\boldsymbol{0}$ to have:

$$\begin{aligned}&[\boldsymbol{\Phi}_{Ly+1}^T(k)\boldsymbol{\Phi}_{Ly+1}(k)+\boldsymbol{\lambda}]\Delta\boldsymbol{u}(k)=\boldsymbol{\Phi}_{Ly+1}^T(k)[(\boldsymbol{y}^*(k+1)-\boldsymbol{y}(k))\\&-\sum_{i=1}^{Ly}\boldsymbol{\Phi}_i(k)\Delta\boldsymbol{y}(k-i+1)-\sum_{i=Ly+2}^{Ly+Lu}\boldsymbol{\Phi}_i(k)\Delta\boldsymbol{u}(k-i+1)]\end{aligned} \tag{7}$$

*Remark 2:* If $\boldsymbol{\lambda}=\boldsymbol{0}$, (7) may be the optimal solution for the tracking error control. It was also shown in [6] for SISO systems.

### *B. Stability Analysis of FFDL-MFAC*

This section provides the performance analysis of MFAPC.

We define

$$\boldsymbol{\phi}_{Ly}(z^{-1})=\boldsymbol{\Phi}_1(k)+\cdots+\boldsymbol{\Phi}_{Ly}(k)z^{-Ly+1} \tag{8}$$

$$\boldsymbol{\phi}_{Lu}(z^{-1})=\boldsymbol{\Phi}_{Ly+1}(k)+\cdots+\boldsymbol{\Phi}_{Ly+Lu}(k)z^{-Lu+1} \tag{9}$$

where $z^{-1}$ is the backward shift operator.

Then (5) at the time $k$ is rewritten as

$$\Delta\boldsymbol{y}(k+1)=\boldsymbol{\phi}_{Ly}(z^{-1})\Delta\boldsymbol{y}(k)+\boldsymbol{\phi}_{Lu}(z^{-1})\Delta\boldsymbol{u}(k) \tag{10}$$

From (7)-(10), we have the following instantaneous closed-loop system equations:

$$\begin{aligned}&\left[\Delta\lambda\left[\boldsymbol{I}-z^{-1}\boldsymbol{\phi}_{Ly}(z^{-1})\right]+\boldsymbol{\phi}_{Lu}(z^{-1})\boldsymbol{\Phi}_{Ly+1}^T(k)\right]\boldsymbol{y}(k+1)\\&=\boldsymbol{\phi}_{Lu}(z^{-1})\boldsymbol{\Phi}_{Ly+1}^T(k)\boldsymbol{y}^*(k+1)\end{aligned} \tag{11}$$

*Theorem 2*: Assume

(1) rank$\left[\boldsymbol{\Phi}_{Ly+1}(k)\right]=M_y$ ($M_u\geq M_y$).

(2) By tuning $\lambda$, we may obtain the inequality:

$$\boldsymbol{T}=\Delta\lambda\left[\boldsymbol{I}-z^{-1}\boldsymbol{\phi}_{Ly}(z^{-1})\right]+\boldsymbol{\phi}_{Lu}(z^{-1})\boldsymbol{\Phi}_{Ly+1}^T(k)\neq\boldsymbol{0},\quad |z|>1 \tag{12}$$

which determines the poles of the system at the time $k$. And (12) guarantees the stability of the system at the time $k$.

Besides, the steady-state error (static error) of the unit-ramp response for the linear system is

$$\begin{aligned}\lim_{k\to\infty}\boldsymbol{e}(k)&=\lim_{\substack{z\to1\\k\to\infty}}\frac{z-1}{z}(\boldsymbol{I}-\boldsymbol{T}^{-1}\boldsymbol{\phi}_{Lu}(z^{-1})\boldsymbol{\Phi}_{Ly+1}^T(k))\frac{T_s z}{(z-1)^2}\\&=\lim_{z\to1}(\boldsymbol{T}^{-1}\lambda\left[\boldsymbol{I}-z^{-1}\boldsymbol{\phi}_{Ly}(z^{-1})\right])T_s\end{aligned} \tag{13}$$

where $T_s$ represents the sample time constant.

The static error of the ramp response of the linear system is proportional to $\lambda$. Furthermore, the steady-state error will be eliminated ( $\lim_{k\to\infty}\boldsymbol{e}(k)=\boldsymbol{0}$ ) when $\boldsymbol{\lambda}=\boldsymbol{0}$ and $M_u\geq M_y$. This noticeable result is distinct from [1]-[5], which have proved that the convergence of tracking error is guaranteed only when $\lambda$ is large enough. Moreover, when system stability can be ensured by setting $\boldsymbol{\lambda}=\boldsymbol{0}$, the static error will be eliminated for a desired trajectory of $k^n\bullet[1,\cdots,1]_{1\times My}^T$ ($0\leq n<\infty$). We omitted the proof. Please refer to [6] for more details.

If rank$\left[\boldsymbol{\Phi}_{Ly+1}(k)\right]=M_u$ ( $M_u<M_y$), we will have the similar stability proof. However, the static error can hardly be guaranteed to be zero.

*Remark 3*: a) If the system exhibits strong nonlinearity, the obtained $\boldsymbol{\phi}_L(k)$ may vary significantly from time $k$ to $k$+1, often resulting in undesirable system performance. Therefore, we suggest applying an iterative MFAC controller as described in [8]. The controller will be

$$
\begin{aligned}
\Delta\boldsymbol{u}_{(i)}(k) = {} & \left( \boldsymbol{\lambda}_{(i)}(k) + \left[ \frac{\partial \boldsymbol{f}(\boldsymbol{\varphi}_{(i)}(k-1))}{\partial \boldsymbol{u}_{(i)}^T(k-1)} \right] \frac{\partial \boldsymbol{f}(\boldsymbol{\varphi}_{(i)}(k-1))}{\partial \boldsymbol{u}_{(i)}(k-1)} \right)^{-1} \bullet \\
& \left[ \frac{\partial \boldsymbol{f}(\boldsymbol{\varphi}_{(i)}(k-1))}{\partial \boldsymbol{u}_{(i)}^T(k-1)} \right]^T [\boldsymbol{y}^*(k+1) - \boldsymbol{y}_{(i)}(k) \\
& - \sum_{j=1}^{ny+1} \frac{\partial \boldsymbol{f}(\boldsymbol{\varphi}_{(i)}(k-1))}{\partial \boldsymbol{y}_{(i)}^T(k-j)} \Delta\boldsymbol{y}_{(i)}(k-j+1) \\
& - \sum_{j=2}^{nu+1} \frac{\partial \boldsymbol{f}(\boldsymbol{\varphi}_{(i)}(k-1))}{\partial \boldsymbol{u}_{(i)}^T(k-j)} \Delta\boldsymbol{u}_{(i)}(k-j+1)]
\end{aligned}
\tag{14}
$$

where ($i$) denotes the iteration count before the control inputs are sent to the system at time $k$+1; To save space, $\boldsymbol{y}_{(i)}(k)$ is a shorthand for $\boldsymbol{y}(k+i\,|\,k)$, $\Delta\boldsymbol{u}_{(i)}(k-j+1)$ for $\Delta\boldsymbol{u}(k-j+1+i\,|\,k)$, $\dfrac{\partial \boldsymbol{f}(\boldsymbol{\varphi}_{(i)}(k-1))}{\partial \boldsymbol{u}_{(i)}^T(k-j))}$ for $\dfrac{\partial \boldsymbol{f}(\boldsymbol{\varphi}(k-1+i\,|\,k))}{\partial \boldsymbol{u}^T(k-j+i\,|\,k))}$, and so on.

b) On the other hand, if the function $\boldsymbol{f}(\cdots)$ has derivatives of all orders at any operating point, the system model (1) can be described as (4) by the Appendix.

The corresponding PJM matrix is

$$
\begin{aligned}
\boldsymbol{\phi}_L^T(k) &= [\boldsymbol{\Phi}_1(k), \cdots, \boldsymbol{\Phi}_{Ly+Lu}(k)] \\
&= [\frac{\partial \boldsymbol{f}(\boldsymbol{\varphi}(k-1))}{\partial \boldsymbol{y}^T(k-1)} + \boldsymbol{\varepsilon}_1(k), \cdots, \frac{\partial \boldsymbol{f}(\boldsymbol{\varphi}(k-1))}{\partial \boldsymbol{y}^T(k-n_y-1)} + \boldsymbol{\varepsilon}_{Ly}(k), \\
&\quad \frac{\partial \boldsymbol{f}(\boldsymbol{\varphi}(k-1))}{\partial \boldsymbol{u}^T(k-1)} + \boldsymbol{\varepsilon}_{Ly+1}(k), \cdots, \frac{\partial \boldsymbol{f}(\boldsymbol{\varphi}(k-1))}{\partial \boldsymbol{u}^T(k-n_u-1)} + \boldsymbol{\varepsilon}_{Ly+Lu}(k)]
\end{aligned}
\tag{15}
$$

where

$$
\boldsymbol{\varepsilon}_i(k) = \frac{1}{2!} \begin{bmatrix} \Delta\boldsymbol{y}^T(k-i+1) \dfrac{\partial^2 f_1(\boldsymbol{\varphi}(k-1))}{\partial[\boldsymbol{y}(k-i)]\partial[\boldsymbol{y}^T(k-i)]} \\ \vdots \\ \Delta\boldsymbol{y}^T(k-i+1) \dfrac{\partial^2 f_{My}(\boldsymbol{\varphi}(k-1))}{\partial[\boldsymbol{y}(k-i)]\partial[\boldsymbol{y}^T(k-i)]} \end{bmatrix} + \cdots \tag{16}
$$

$$
\boldsymbol{\varepsilon}_{Ly+j}(k) = \frac{1}{2!} \begin{bmatrix} \Delta\boldsymbol{u}^T(k-j+1) \dfrac{\partial^2 f_1(\boldsymbol{\varphi}(k-1))}{\partial[\boldsymbol{u}(k-j)]\partial[\boldsymbol{u}^T(k-j)]} \\ \vdots \\ \Delta\boldsymbol{u}^T(k-j+1) \dfrac{\partial^2 f_{My}(\boldsymbol{\varphi}(k-1))}{\partial[\boldsymbol{u}(k-j)]\partial[\boldsymbol{u}^T(k-j)]} \end{bmatrix} + \cdots \tag{17}
$$

$i$=1,$\cdots$,$L_y$ and $j$=1,$\cdots$,$L_u$.

Further, if $\dfrac{\partial \boldsymbol{f}(\boldsymbol{\varphi}(k-1))}{\partial \boldsymbol{u}^T(k-1)} \neq \boldsymbol{0}$, we may obtain the control law by minimizing the cost function (18) or the cost function (19) subject to constraint $\boldsymbol{G}(\boldsymbol{u}(k)) \le \boldsymbol{u}_M(k)$.

$$
\begin{aligned}
J &= \left[ \boldsymbol{y}^*(k+1) - \boldsymbol{y}(k+1) \right]^T \left[ \boldsymbol{y}^*(k+1) - \boldsymbol{y}(k+1) \right] + \Delta\boldsymbol{u}^T(k)\boldsymbol{\lambda}\Delta\boldsymbol{u}(k) \\
&= \Bigg\| \boldsymbol{y}^*(k+1) - \boldsymbol{y}(k) - \sum_{i=Ly+1}^{Ly+Lu} \boldsymbol{\Phi}_i(k)\Delta\boldsymbol{u}(k+L_y-i+1) \\
&\quad - \sum_{i=1}^{Ly} \boldsymbol{\Phi}_i(k)\Delta\boldsymbol{y}(k-i+1) \Bigg\|_2 + \Delta\boldsymbol{u}^T(k)\boldsymbol{\lambda}\Delta\boldsymbol{u}(k)
\end{aligned}
\tag{18}
$$

$$
\begin{aligned}
\min_{\boldsymbol{u}(k) \text{ s.t. } \boldsymbol{G}(\boldsymbol{u}(k)) \le \boldsymbol{u}_M(k)} J &= \left[ \boldsymbol{y}^*(k+1) - \boldsymbol{y}(k+1) \right]^T \left[ \boldsymbol{y}^*(k+1) - \boldsymbol{y}(k+1) \right] \\
&\quad + \Delta\boldsymbol{u}^T(k)\boldsymbol{\lambda}\Delta\boldsymbol{u}(k)
\end{aligned}
\tag{19}
$$

In applications, we may simplify the controller design process through an approximation (20) for the design of the controller (7).

$$
\begin{aligned}
\boldsymbol{\phi}_L^T(k) = {} & [\frac{\partial \boldsymbol{f}(\boldsymbol{\varphi}(k-1))}{\partial \boldsymbol{y}^T(k-1)}, \cdots, \frac{\partial \boldsymbol{f}(\boldsymbol{\varphi}(k-1))}{\partial \boldsymbol{y}^T(k-n_y-1)}, \frac{\partial \boldsymbol{f}(\boldsymbol{\varphi}(k-1))}{\partial \boldsymbol{u}^T(k-1)}, \\
& \cdots, \frac{\partial \boldsymbol{f}(\boldsymbol{\varphi}(k-1))}{\partial \boldsymbol{u}^T(k-n_u-1)}]
\end{aligned}
\tag{20}
$$

## IV. SIMULATION

Example 1: In this example, we assume that the offline model is known for making comparisons among the tree controllers designed through minimizing (18) or (19), and the MFAC controller (7) with roughly calculated PJM by (20). The model is given as the following nonlinear system:

$$
\begin{aligned}
y_1(k+1) &= -0.1y_1^3(k) + 0.2y_2^2(k) + u_1(k) + u_2^2(k) + u_1^3(k-1) + 2u_1^4(k-1) \\
y_2(k+1) &= -0.1y_1^2(k) + 0.2y_2^3(k) + u_1^2(k) + 0.8u_2(k) + u_1^3(k-1) + u_2^3(k-1)
\end{aligned}
\tag{21}
$$

The desired output trajectories are

$$
\begin{aligned}
y_1^*(k) &= 0.3\sin(k/40) - 0.2\cos(k/20) && 1 \le k \le 400 \\
y_2^*(k) &= 0.2\sin(k/10) + 0.3\sin(k/30) && 1 \le k \le 400 \\
y_1^*(k) &= -y_2^*(k) = 0.2 \times (-1)^{round(k/50)} && 401 \le k \le 800
\end{aligned}
\tag{22}
$$

The initial values are $\boldsymbol{y}(1) = \boldsymbol{y}(2) = \boldsymbol{y}(3) = \boldsymbol{u}(1) = \boldsymbol{u}(2) = \boldsymbol{0}$, $\boldsymbol{\phi}_L^T(1) = \boldsymbol{\phi}_L^T(2) = 0.01 \bullet ones(2,6)$. The controller structure is $L_y$=$n_y$+1=1, $L_u$=$n_u$+1=2. We choose $\boldsymbol{\lambda}$=0.2$\boldsymbol{I}$. The controller MFAC 1 is designed by minimizing the quartic equation (18), shown as

$$
\begin{aligned}
J &= \left[ \boldsymbol{y}^*(k+1) - \boldsymbol{y}(k+1) \right]^T \left[ \boldsymbol{y}^*(k+1) - \boldsymbol{y}(k+1) \right] + \Delta\boldsymbol{u}^T(k)\boldsymbol{\lambda}\Delta\boldsymbol{u}(k) \\
&= \Delta\boldsymbol{u}^T(k)\boldsymbol{\lambda}\Delta\boldsymbol{u}(k) + \Big\| \boldsymbol{y}^*(k+1) - \boldsymbol{y}(k) - \boldsymbol{\Phi}_1(k)\Delta\boldsymbol{y}(k) - \boldsymbol{\Phi}_2(k)\Delta\boldsymbol{u}(k) \\
&\quad - \boldsymbol{\Phi}_3(k)\Delta\boldsymbol{u}(k-1) \Big\|_2
\end{aligned}
\tag{23}
$$

with $\boldsymbol{\phi}_L^T(k) = [\boldsymbol{\Phi}_1(k), \cdots, \boldsymbol{\Phi}_{Ly+Lu}(k)]$ calculated by (15)-(17).

Similarly, the controller MFAC 2 is designed by minimizing (20) subject to constraints $-0.3 \le u_1(k) \le 0.1$ and $-0.5 \le u_2(k) \le 0.5$.

We apply the controller (7) named MFAC 3. The PJM is roughly calculated by (20). The outputs of the system are shown in Fig. 1 and Fig. 2, respectively. The outputs of three controllers are shown in Fig. 3. Fig. 4 shows the calculated elements in PJM for MFAC 1.

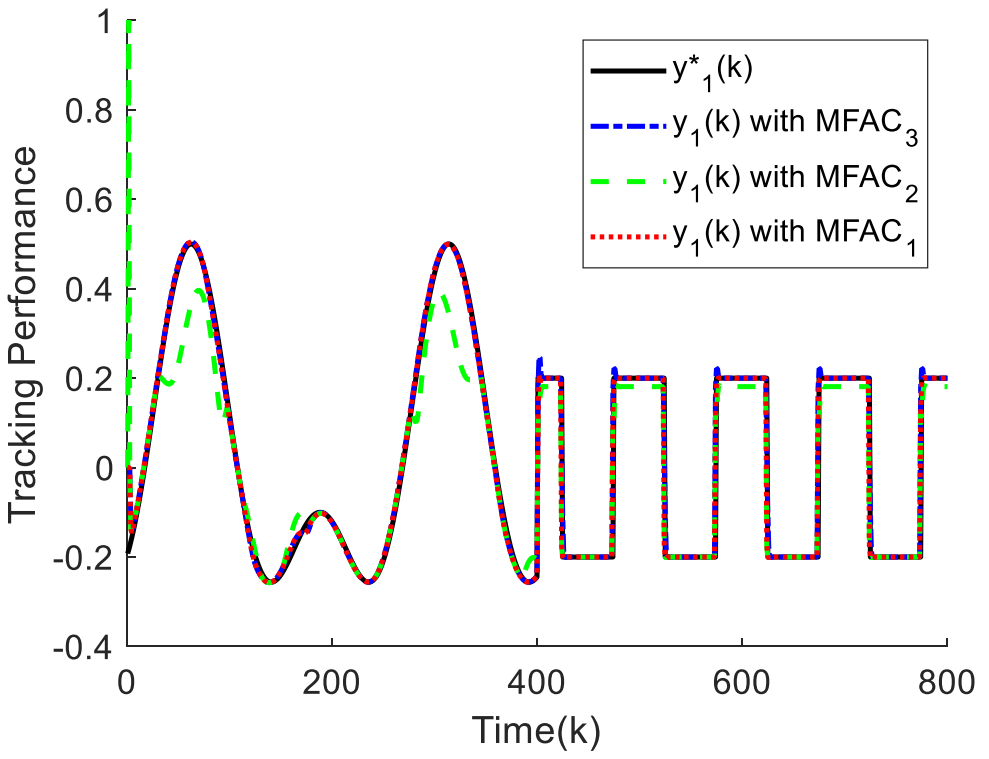

Fig. 1 Tracking performance

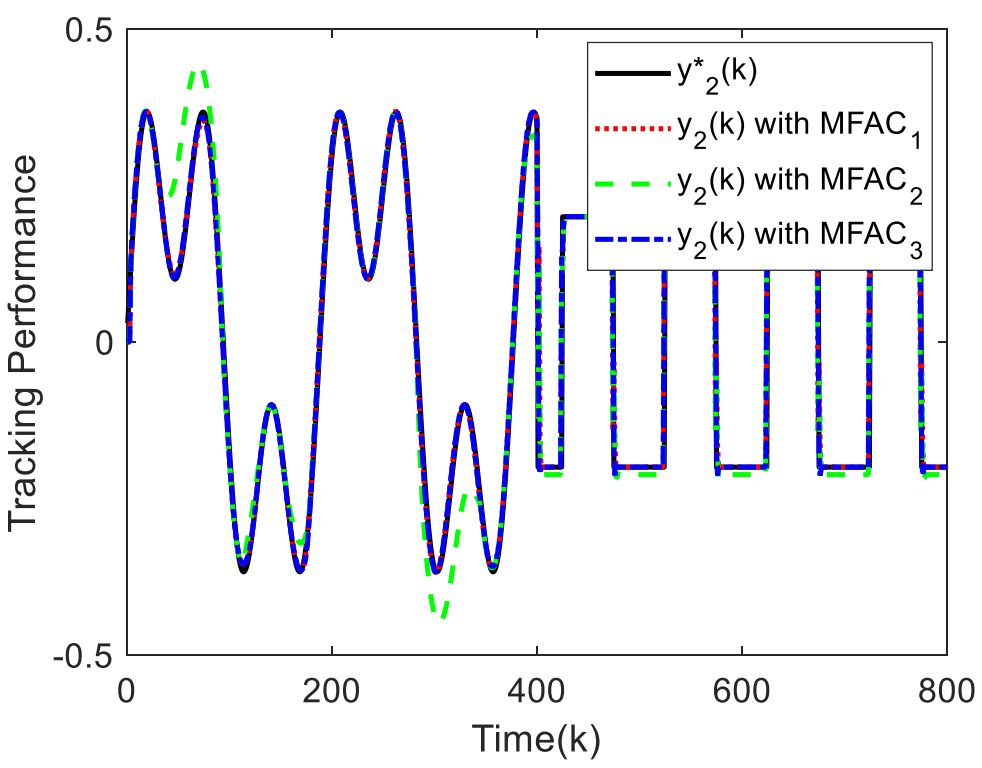

Fig. 2 Tracking performance

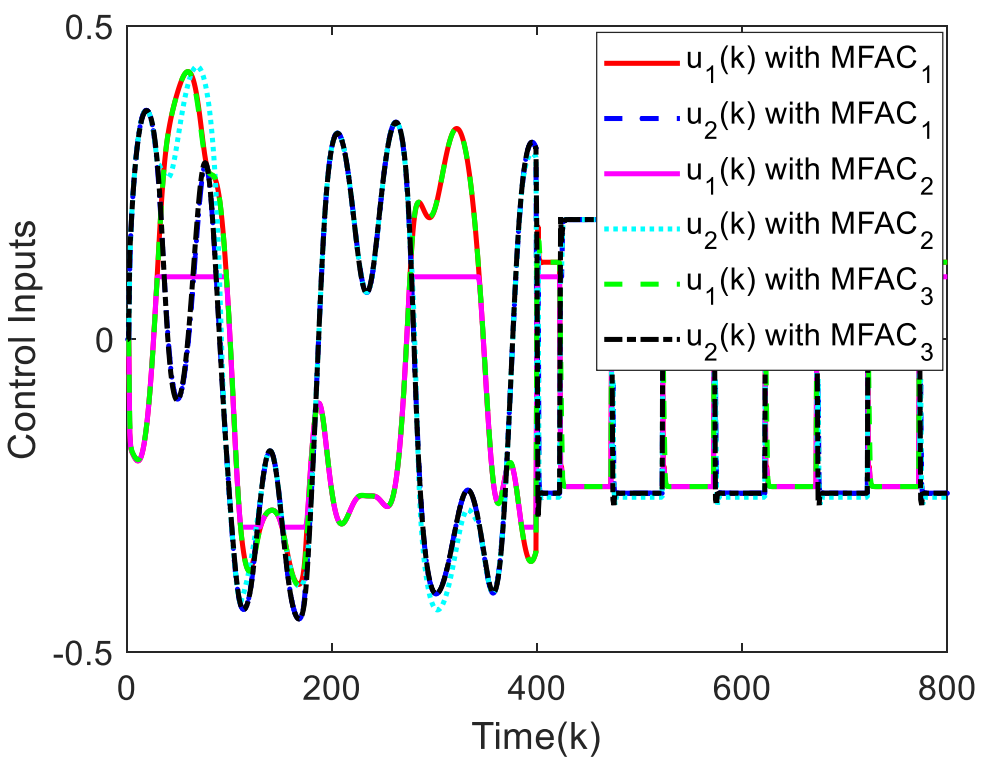

Fig. 3 Control inputs

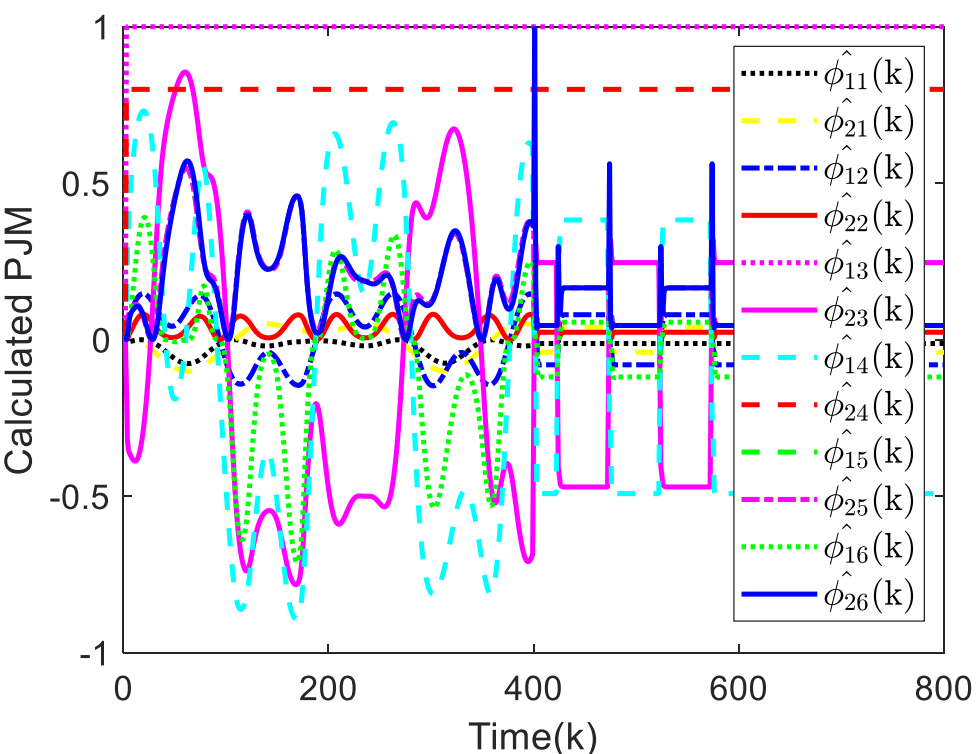

Fig. 4 Elements in calculated PJM

In the following example, we will demonstrate that the utilization of the matrix $\left[\boldsymbol{\Phi}^T(k)\boldsymbol{\Phi}(k)+\lambda(k)\boldsymbol{I}\right]^{-1}\boldsymbol{\Phi}^T(k)$ instead of $\boldsymbol{\Phi}^{-1}(k)$ makes the robot stay stable even near the singularities. For more details, please refer to the inverse kinematic code in ikine.m of MATLAB Robotics Toolbox which is implemented according to [15]-[17]. Additionally, the method in [15] can be considered a special MFAC case when $L_y$=0, $L_u$=1. For more knowledge, please refer to [8].

Example 2: A six-dimensional industrial robot is shown in Fig. 5. Herein, we will design a simple yet useful controller for the industrial robot manipulator. The block diagram of the system is shown in Fig. 6.

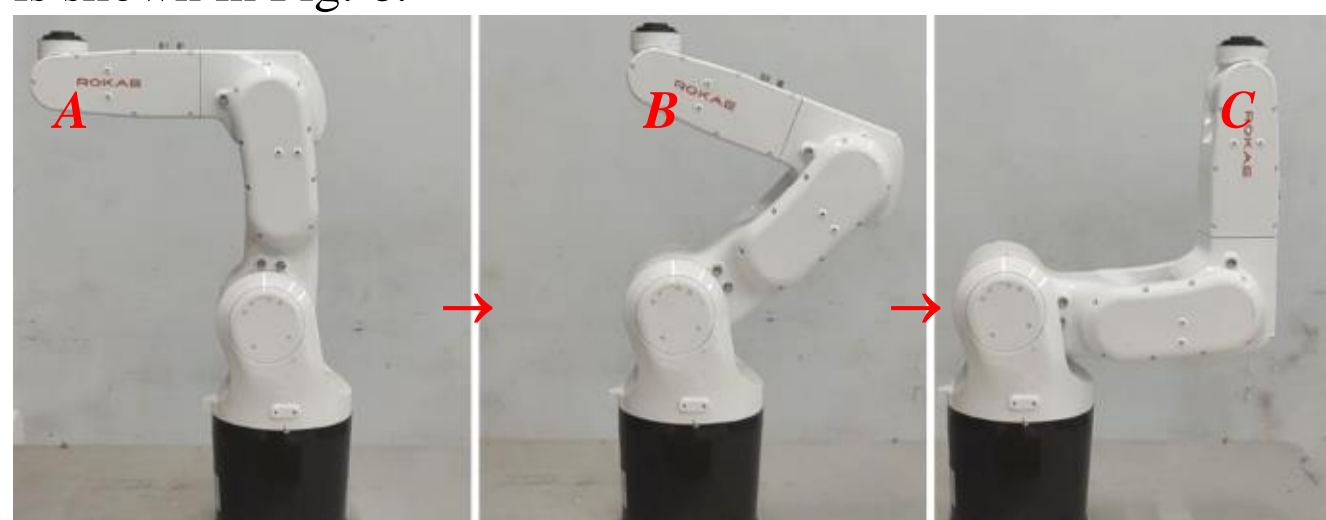

Fig. 5 Robot manipulator

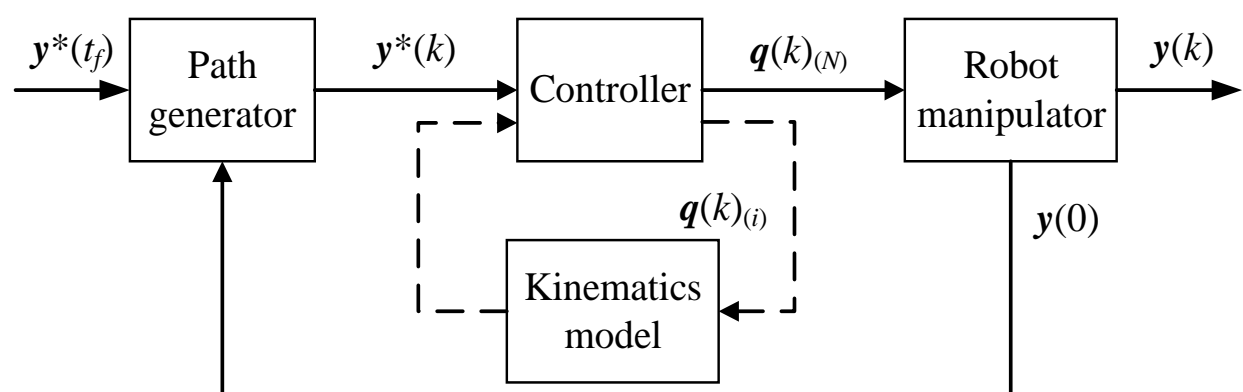

Fig. 6 Block diagram of the system

According to the manual, Table I gives the Denavit-Hartenberg parameters (link twist $\alpha_{i-1}$, link length $a_{i-1}$, link offset $d_i$, and joint angle $q_i$) of the manipulator to calculate the frame in Cartesian space.

TABLE I Denavit-Hartenberg parameters

| $i$ | $\alpha_{i-1}$ /° | $a_{i-1}$ /mm | $d_i$ /mm | $q_i$ |
|---|---|---|---|---|
| Base | 0 | 0 | 380 | 0 |
| 1 | 0 | 0 | 0 | $q_1$ |
| 2 | -90 | 30 | 0 | $q_2$ |
| 3 | 0 | 340 | 0 | $q_3$ |
| 4 | -90 | 35 | 335 | $q_4$ |
| 5 | 90 | 0 | 0 | $q_5$ |
| 6 | -90 | 0 | 0 | $q_6$ |
| Tool | 0 | 0 | 83 | 0 |

**Part A: MFAC controller design for inverse kinematics solutions**

Given the desired path point in the desired trajectory (planned Cartesian-space paths) at path-planning time $k$ is

$$\boldsymbol{y}^*(k)=[x^*(k),y^*(k),z^*(k),\alpha^*(k),\beta^*(k),\gamma^*(k)]^T \tag{24}$$

It is normal to choose the Cartesian-space scheme for the linear spline with parabolic blends path in **Part B** of this example [17].

The corresponding desired frame can be described as

$$\boldsymbol{T}^*(k)=\begin{pmatrix}\boldsymbol{A}^*(k) & \boldsymbol{p}^*(k)\\ \boldsymbol{0}_{1\times3} & 1\end{pmatrix} \tag{25}$$

where

$$\boldsymbol{A}^*(k)=\begin{bmatrix} t_{11}^*(k) & t_{12}^*(k) & t_{13}^*(k) \\ t_{21}^*(k) & t_{22}^*(k) & t_{23}^*(k) \\ t_{31}^*(k) & t_{32}^*(k) & t_{33}^*(k) \end{bmatrix} \tag{26}$$

is the desired rotation matrix and

$$\boldsymbol{p}^*(k)=[x^*(k), y^*(k), z^*(k)]^T \tag{27}$$

is the desired position vector.

$t_{11}^*(k)=c\beta^*(k)c\gamma^*(k)$ , $t_{21}^*(k)=c\beta^*(k)s\gamma^*(k)$ ,

$t_{12}^*(k)=c\gamma^*(k)s\alpha^*(k)s\beta^*(k)-c\alpha^*(k)s\gamma^*(k)$ ,

$t_{13}^*(k)=s\alpha^*(k)s\gamma^*(k)+c\alpha^*(k)c\gamma^*(k)s\beta^*(k)$ ,

$t_{22}^*(k)=c\alpha^*(k)c\gamma^*(k)+s\alpha^*(k)s\gamma^*(k)s\beta^*(k)$ ,

$t_{23}^*(k)=c\alpha^*(k)s\beta^*(k)s\gamma^*(k)-c\gamma^*(k)s\alpha^*(k)$

$t_{31}^*(k)=-s\beta^*(k)$ , $t_{32}^*(k)=c\beta^*(k)s\alpha^*(k)$ ,

$t_{13}^*(k)=c\alpha^*(k)c\beta^*(k)$ ,

where $c\bullet$ is shorthand for $\cos(\bullet)$ , $s\bullet$ for $\sin(\bullet)$ , and so on. The calculated robot frame at the *i*-th iteration for solving the inverse kinematics concerning the desired path point $\boldsymbol{y}^*(k+1)$ is described by

$$\boldsymbol{T}(k)_{(i)}=\begin{bmatrix} t(k)_{11(i)} & t(k)_{12(i)} & t(k)_{13(i)} & x(k)_{(i)} \\ t(k)_{21(i)} & t(k)_{22(i)} & t(k)_{23(i)} & y(k)_{(i)} \\ t(k)_{31(i)} & t(k)_{32(i)} & t(k)_{33(i)} & z(k)_{(i)} \\ 0 & 0 & 0 & 1 \end{bmatrix} \tag{28}$$

Hence, the corresponding orientation matrix is

$$\boldsymbol{A}(k)_{(i)}=\begin{bmatrix} t(k)_{11(i)} & t(k)_{12(i)} & t(k)_{13(i)} \\ t(k)_{21(i)} & t(k)_{22(i)} & t(k)_{23(i)} \\ t(k)_{31(i)} & t(k)_{32(i)} & t(k)_{33(i)} \end{bmatrix} \tag{29}$$

Then the desired orientation matrix relative to the calculated orientation matrix of the manipulator at the *i*-th iteration is obtained

$$\boldsymbol{D}(k)_{(i)}=\begin{bmatrix} d(k)_{11(i)} & d(k)_{12(i)} & d(k)_{13(i)} \\ d(k)_{21(i)} & d(k)_{22(i)} & d(k)_{23(i)} \\ d(k)_{31(i)} & d(k)_{32(i)} & d(k)_{33(i)} \end{bmatrix} = \boldsymbol{A}^*(k+1)\left(\boldsymbol{A}(k)_{(i)}\right)^{-1} \tag{30}$$

Now, we convert the orientation matrix into an equivalent angle-axis representation to calculate the tracking error of the Euler angles vector. The angle is

$$\theta(k)_{(i)}=\arccos\left(\frac{d(k)_{11(i)}+d(k)_{22(i)}+d(k)_{33(i)}-1}{2}\right) \tag{31}$$

The equivalent axis of a finite rotation is

$$\hat{\boldsymbol{K}}(k)_{(i)}=\begin{pmatrix} d(k)_{32(i)}-d(k)_{23(i)} \\ d(k)_{13(i)}-d(k)_{31(i)} \\ d(k)_{21(i)}-d(k)_{12(i)} \end{pmatrix} \Big/ 2\sin\theta(k)_{(i)} \tag{32}$$

Then the relative rotational Euler angles vector is calculated by

$$\begin{bmatrix} \alpha^*(k+1)-\alpha(k)_{(i)} \\ \beta^*(k+1)-\beta(k)_{(i)} \\ \gamma^*(k+1)-\gamma(k)_{(i)} \end{bmatrix} = \hat{\boldsymbol{K}}(k)_{(i)}\theta(k)_{(i)} \tag{33}$$

The control law is

$$\begin{aligned} &\Delta\boldsymbol{q}(k)_{(i)} \\ &=[\boldsymbol{\Phi}^T(k)_{(i)}\boldsymbol{\Phi}(k)_{(i)}+\boldsymbol{\lambda}(k)_{(i)}]^{-1}\boldsymbol{\Phi}^T(k)_{(i)}[(\boldsymbol{y}^*(k+1)-\boldsymbol{y}(k)_{(i)})] \\ &=[\boldsymbol{\Phi}^T(k)_{(i)}\boldsymbol{\Phi}(k)_{(i)}+\boldsymbol{\lambda}(k)_{(i)}]^{-1}\boldsymbol{\Phi}^T(k)_{(i)}[x^*(k+1)-x(k)_{(i)}, \\ &\quad y^*(k+1)-y(k)_{(i)}, z^*(k+1)-z(k)_{(i)}, \alpha^*(k+1)-\alpha(k)_{(i)}, \\ &\quad \beta^*(k+1)-\beta(k)_{(i)}, \gamma^*(k+1)-\gamma(k)_{(i)}]^T \end{aligned} \tag{34}$$

We adjust the controller parameters by the following lookup table

$$\boldsymbol{\lambda}(k)_{(i+1)}=\begin{cases} 0\boldsymbol{I} & \text{if} \quad \mathrm{cond}(\boldsymbol{\Phi}(k)_{(i)})<5000 \\ 0.05\boldsymbol{I} & \text{if} \quad 5000\le\mathrm{cond}(\boldsymbol{\Phi}(k)_{(i)})<20000 \\ 0.1\boldsymbol{I} & \text{if} \quad \mathrm{cond}(\boldsymbol{\Phi}(k)_{(i)})\ge 20000 \end{cases} \tag{35}$$

The joint angle vector of the manipulator at the *i*-th iteration is

$$\boldsymbol{q}(k)_{(i)}=\boldsymbol{q}(k)_{(i\text{-}1)}+\Delta\boldsymbol{q}(k)_{(i)} \tag{36}$$

According to the robot kinematics, we can transform $\boldsymbol{q}(k)_{(i)}$ into

$$\boldsymbol{y}(k)_{(i+1)}=[x(k)_{(i+1)}, y(k)_{(i+1)}, z(k)_{(i+1)}, \alpha(k)_{(i+1)}, \beta(k)_{(i+1)}, \gamma(k)_{(i+1)}]^T \tag{37}$$

which can be directly transformed into (28) for the (*i*+1)-th iteration. The Jacobian matrix $\boldsymbol{\Phi}(k)_{(i+1)}$ is only decided by $\boldsymbol{q}(k)_{(i)}$ . Then (28)-(37) forms a closed loop to calculate the inverse kinematics solution for the desired path point $\boldsymbol{y}^*(k+1)$ .

In this example, the maximum number of iterations is limited to 30.

**Part B: Path generator**

Herein, the generation of Cartesian-space paths is designed for the linear spline with quantic blends path.

We plan the straight-line path that begins with the initial frame of the manipulator $[x(0), y(0), z(0), \alpha(0), \beta(0), \gamma(0)]^T$ and ends at the goal frame $[x^*(t_f), y^*(t_f), z^*(t_f), \alpha^*(t_f), \beta^*(t_f), \gamma^*(t_f)]^T$ . Thus, the initial position vector is written as $\boldsymbol{p}(0)=[x(0), y(0), z(0)]^T$ and the goal position vector is written as $\boldsymbol{p}^*(t_f)=[x^*(t_f), y^*(t_f), z^*(t_f)]^T$ .

A quantic polynomial is used for the Cartesian-straight-line-motion scheme

$$S_1(t)=a_0+a_1t+a_2t^2+a_3t^3+a_4t^4+a_5t^5, \qquad 0\le t\le t_f \tag{38}$$

The coefficients $a_i$ ($i$=1,⋯,5) are specified by the constraints: initial velocity $\dot{\boldsymbol{p}}(0)$ , initial acceleration $\ddot{\boldsymbol{p}}(0)$ , goal position $S_1(t_f)$ ( $\boldsymbol{p}^*(t_f)$ ), goal velocity $\dot{\boldsymbol{p}}^*(t_f)$ and goal acceleration $\ddot{\boldsymbol{p}}^*(t_f)$ at the final time $t_f$ .The solution of $a_i$ is

$$a_0=0,\ a_1=\left\|\dot{\boldsymbol{p}}(0)\right\|_2,\ a_2=\left\|\ddot{\boldsymbol{p}}(0)\right\|_2/2$$

$$a_3=\frac{20S_1(t_f)-(8\left\|\dot{\boldsymbol{p}}^*(t_f)\right\|_2+12\left\|\dot{\boldsymbol{p}}(0)\right\|_2)t_f-(3\left\|\ddot{\boldsymbol{p}}(0)\right\|_2-\left\|\ddot{\boldsymbol{p}}^*(t_f)\right\|_2)t_f^2}{2t_f^3}$$

$$a_4=\frac{-30S_1(t_f)-(14\left\|\dot{\boldsymbol{p}}^*(t_f)\right\|_2+16\left\|\dot{\boldsymbol{p}}(0)\right\|_2)t_f+(3\left\|\ddot{\boldsymbol{p}}(0)\right\|_2-2\left\|\ddot{\boldsymbol{p}}^*(t_f)\right\|_2)t_f^2}{2t_f^4}$$

$$a_5 = \frac{12S_1(t_f) - 6(\|\dot{\boldsymbol{p}}^*(t_f)\|_2 + \|\dot{\boldsymbol{p}}(0)\|_2)t_f - (\|\ddot{\boldsymbol{p}}(0)\|_2 - \|\ddot{\boldsymbol{p}}^*(t_f)\|_2)t_f^2}{2t_f^5} \tag{39}$$

Then the positional trajectory at path-planning time $k$, i.e. (27), is calculated by

$$\boldsymbol{p}^*(k) = \boldsymbol{p}(0) + \frac{\boldsymbol{p}^*(t_f) - \boldsymbol{p}(0)}{\|\boldsymbol{p}^*(t_f) - \boldsymbol{p}(0)\|_2} \bullet S_1(t) + \boldsymbol{p}(0) \tag{40}$$

Where $t = kT_0$ [?] $T_0$ is the implementation period of the manipulator.

We know that the initial X-Y-Z Euler angle vector is $\boldsymbol{\Theta}(0) = [\alpha(0), \beta(0), \gamma(0)]^T$ and the goal X-Y-Z Euler angle vector is $\boldsymbol{\Theta}(t_f) = [\alpha^*(t_f), \beta^*(t_f), \gamma^*(t_f)]^T$.

We convert the goal X-Y-Z Euler angle vector into unit quaternion (Euler parameters):

$$\boldsymbol{\Pi}^*(t_f) = [\varepsilon_1(t_f) \quad \varepsilon_2(t_f) \quad \varepsilon_3(t_f) \quad \varepsilon_4(t_f)]^T$$
$$= \begin{bmatrix} s\frac{\alpha^*(t_f)}{2} c\frac{\beta^*(t_f)}{2} c\frac{\gamma^*(t_f)}{2} - c\frac{\alpha^*(t_f)}{2} s\frac{\beta^*(t_f)}{2} s\frac{\gamma^*(t_f)}{2} \\ c\frac{\alpha^*(t_f)}{2} s\frac{\beta^*(t_f)}{2} c\frac{\gamma^*(t_f)}{2} + s\frac{\alpha^*(t_f)}{2} c\frac{\beta^*(t_f)}{2} s\frac{\gamma^*(t_f)}{2} \\ c\frac{\alpha^*(t_f)}{2} c\frac{\beta^*(t_f)}{2} s\frac{\gamma^*(t_f)}{2} + s\frac{\alpha^*(t_f)}{2} s\frac{\beta^*(t_f)}{2} c\frac{\gamma^*(t_f)}{2} \\ c\frac{\alpha^*(t_f)}{2} c\frac{\beta^*(t_f)}{2} c\frac{\gamma^*(t_f)}{2} + s\frac{\alpha^*(t_f)}{2} s\frac{\beta^*(t_f)}{2} s\frac{\gamma^*(t_f)}{2} \end{bmatrix} \tag{41}$$

Similarly, we can convert $\boldsymbol{\Theta}(0)$ into quaternion $\boldsymbol{\Pi}(0) = [\varepsilon_1(0), \varepsilon_2(0), \varepsilon_3(0), \varepsilon_4(0)]^T$.

The length of the path is

$$S_2(t_f) = \left\| 2\boldsymbol{\Pi}(0)\log\left(\boldsymbol{\Pi}^{-1}(0)\boldsymbol{\Pi}^*(t_f)\right)\boldsymbol{\Pi}^{-1}(0) \right\|_2 \tag{42}$$

Similarly, the quantic polynomial is used for the Cartesian-straight-line-motion scheme

$$S_2(t) = b_0 + b_1 t + b_2 t^2 + b_3 t^3 + b_4 t^4 + b_5 t^5, \qquad t \le t_f \tag{43}$$

Similar to (39), $b_i$ ($i$=1,⋯,5) are specified by the constraints: $\boldsymbol{\Pi}(0)$, $\dot{\boldsymbol{\Pi}}(0)$, $\ddot{\boldsymbol{\Pi}}(0)$, $S(t_f)$, $\dot{\boldsymbol{\Pi}}^*(t_f)$ and $\ddot{\boldsymbol{\Pi}}^*(t_f)$.

Define

$$\begin{bmatrix} \bar{\varepsilon}_1 & \bar{\varepsilon}_2 & \bar{\varepsilon}_3 & \bar{\varepsilon}_4 \end{bmatrix}^T = \boldsymbol{\Pi}^{-1}(0)\boldsymbol{\Pi}^*(t_f) \tag{44}$$

We obtain the quaternion argument

$$a = \arccos \bar{\varepsilon}_4 \tag{45}$$

Then we have the desired orientation at path-planning time $k$

$$\boldsymbol{\Pi}(k) = \left[\varepsilon_1(k) \quad \varepsilon_2(k) \quad \varepsilon_3(k) \quad \varepsilon_4(k)\right]^T$$
$$= \boldsymbol{\Pi}(0)\left[\boldsymbol{\Pi}^{-1}(0)\boldsymbol{\Pi}^*(t_f)\right]^\tau, \qquad \tau = \frac{S_2(t)}{S_2(t_f)} \tag{46}$$

where

$$\left[\boldsymbol{\Pi}^{-1}(0)\boldsymbol{\Pi}^*(t_f)\right]^\tau = \exp(\tau \log\left(\boldsymbol{\Pi}^{-1}(0)\boldsymbol{\Pi}^*(t_f)\right))$$
$$= \exp\left([0, a\tau\left(\bar{\varepsilon}_1/sa \quad \bar{\varepsilon}_2/sa \quad \bar{\varepsilon}_3/sa\right)\right)$$
$$= [\bar{\varepsilon}_1\, s(a\,S_2(t)/S_2(t_f))/sa, \bar{\varepsilon}_2\, s(a\,S_2(t)/S_2(t_f))/sa,$$
$$\bar{\varepsilon}_3\, s(a\,S_2(t)/S_2(t_f))/sa), c(a\,S_2(t)/S_2(t_f))]^T, \qquad t = kT_0 \tag{47}$$

and exp(•) represents exponential operation; log(•) represents the logarithmic operation. We convert (46) into an X-Y-Z Euler angle vector

$$\boldsymbol{\Theta}^*(k) = [\alpha^*(k), \beta^*(k), \gamma^*(k)]^T$$
$$= \begin{bmatrix} \arctan\left(2(\varepsilon_4(k)\varepsilon_1(k) + \varepsilon_2(k)\varepsilon_3(k))/(1 - 2(\varepsilon_1^2(k) + \varepsilon_2^2(k))\right) \\ \arcsin\left(2(\varepsilon_4(k)\varepsilon_2(k) - \varepsilon_3(k)\varepsilon_1(k)\right) \\ \arctan\left(2(\varepsilon_4(k)\varepsilon_3(k) + \varepsilon_1(k)\varepsilon_2(k))/(1 - 2(\varepsilon_2^2(k) + \varepsilon_3^2(k))\right) \end{bmatrix} \tag{48}$$

which can be transformed into (26) directly. We finish the design of the desired trajectory (24).

**Part C: Test MFAC control law**

Herein, The MFAC in Part A is tested by tracking the trajectory defined in the task space. The beginning joint angle vector of the manipulator is $\boldsymbol{A}\,[-\pi/2, 0, 0, 0, -\pi/2, 0]^T$. The path generator in Part B is used to plan a straight-line path in Cartesian space from $\boldsymbol{A}$ to $\boldsymbol{C}$ $[\pi/2, 0, 0, 0, -\pi/2, 0]^T$. Consequently, there are two singular points $\boldsymbol{B}$ and $\boldsymbol{C}$ in the trajectory. The manipulator frames $\boldsymbol{A}$, $\boldsymbol{B}$ and $\boldsymbol{C}$ are marked in Fig. 5. Fig. 7 shows the tracking performance of the manipulator and Fig. 8 shows the corresponding tracking error.

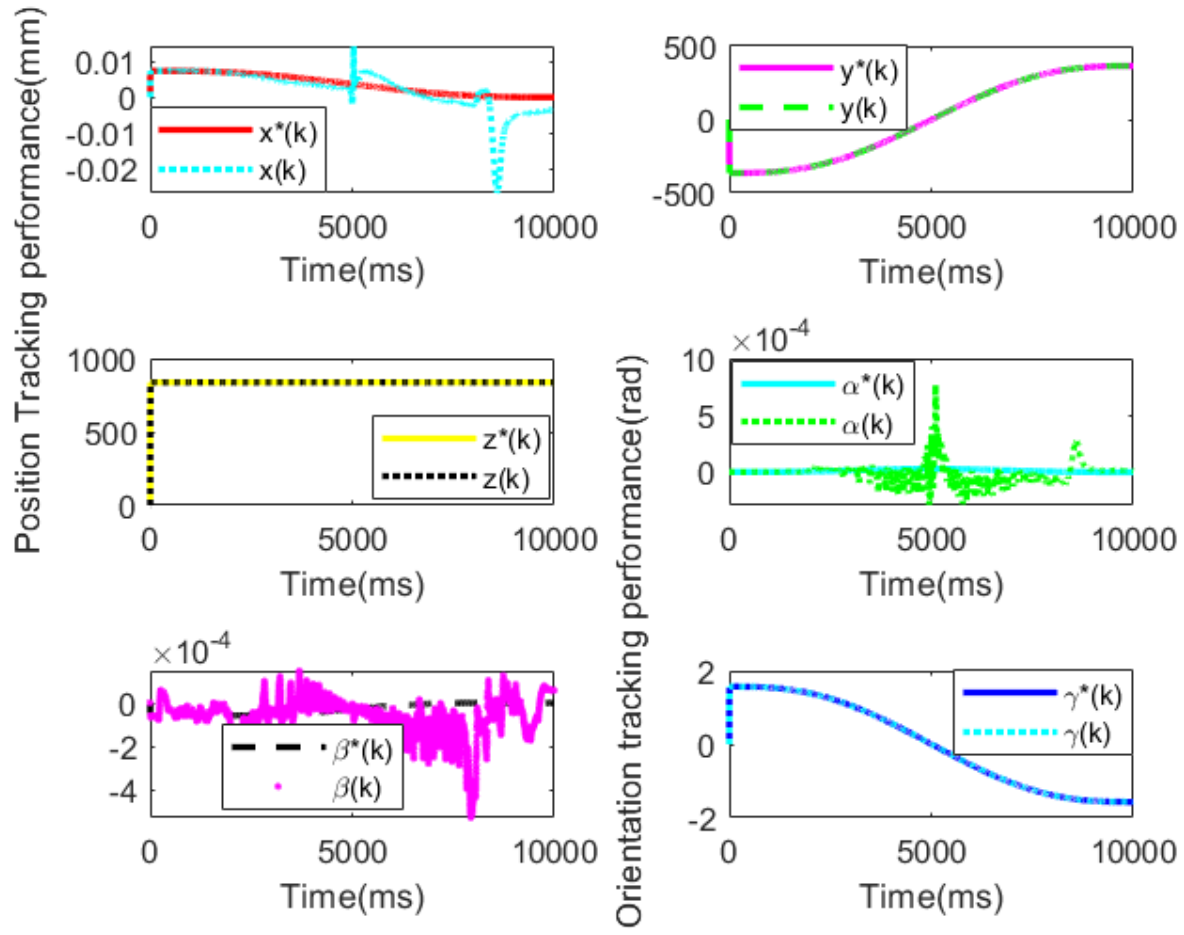

Fig. 7 Tracking performance

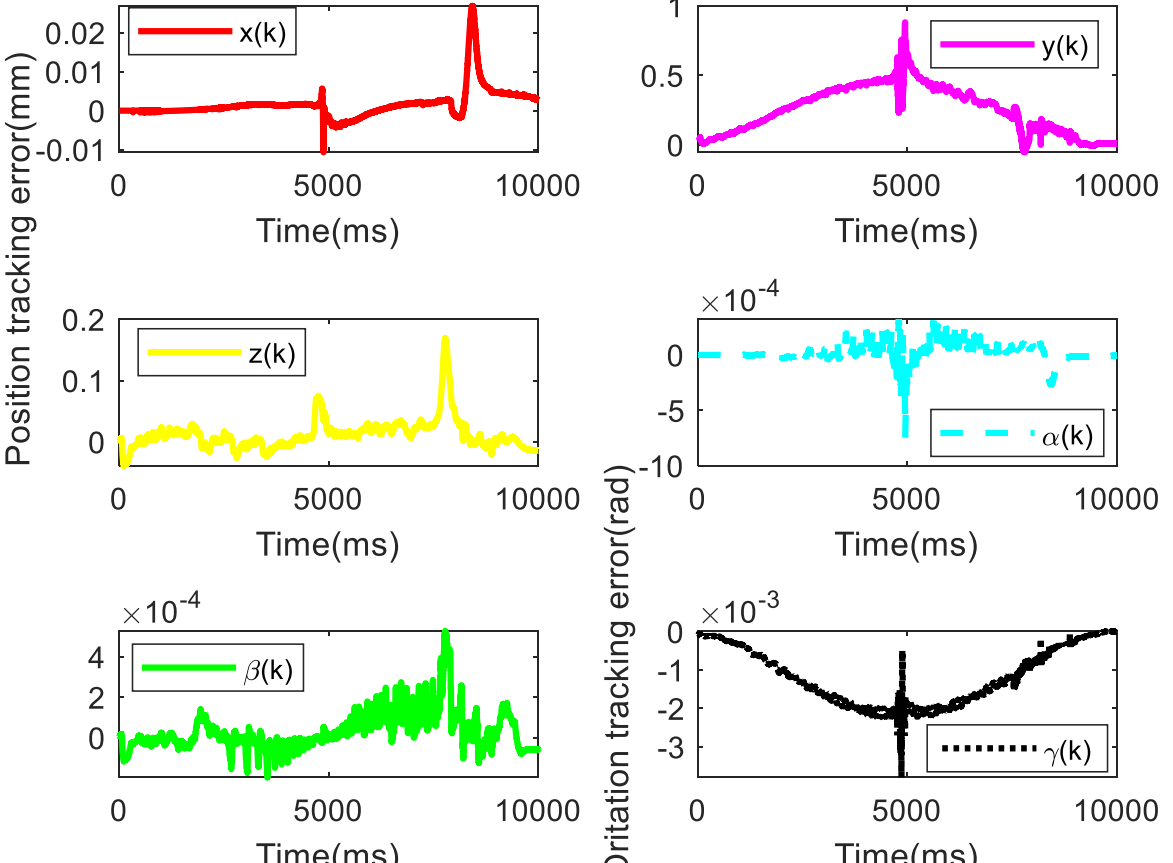

Fig. 8 Tracking error

Fig. 9 shows the measured joint angles. Fig. 10 shows a part of the elements in $\hat{\boldsymbol{\Phi}}(k)$. Fig. 11 shows the condition number and Fig. 12 shows the controller parameter $\lambda$. Fig. 13 shows the iteration count.

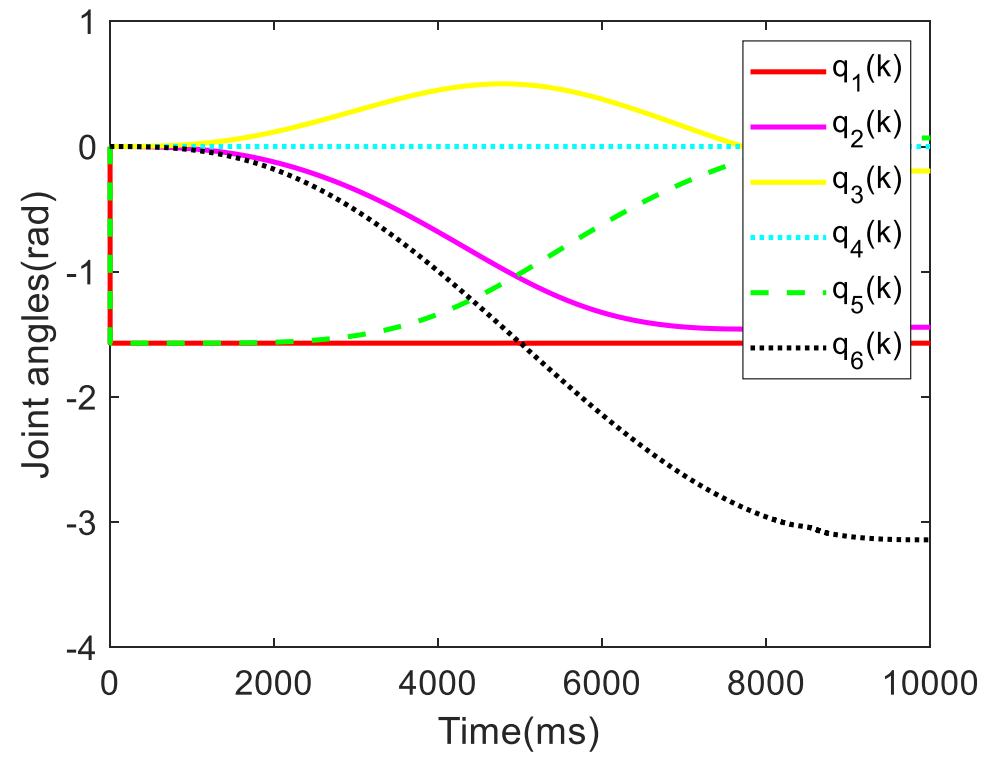

Fig. 9 Joint angles

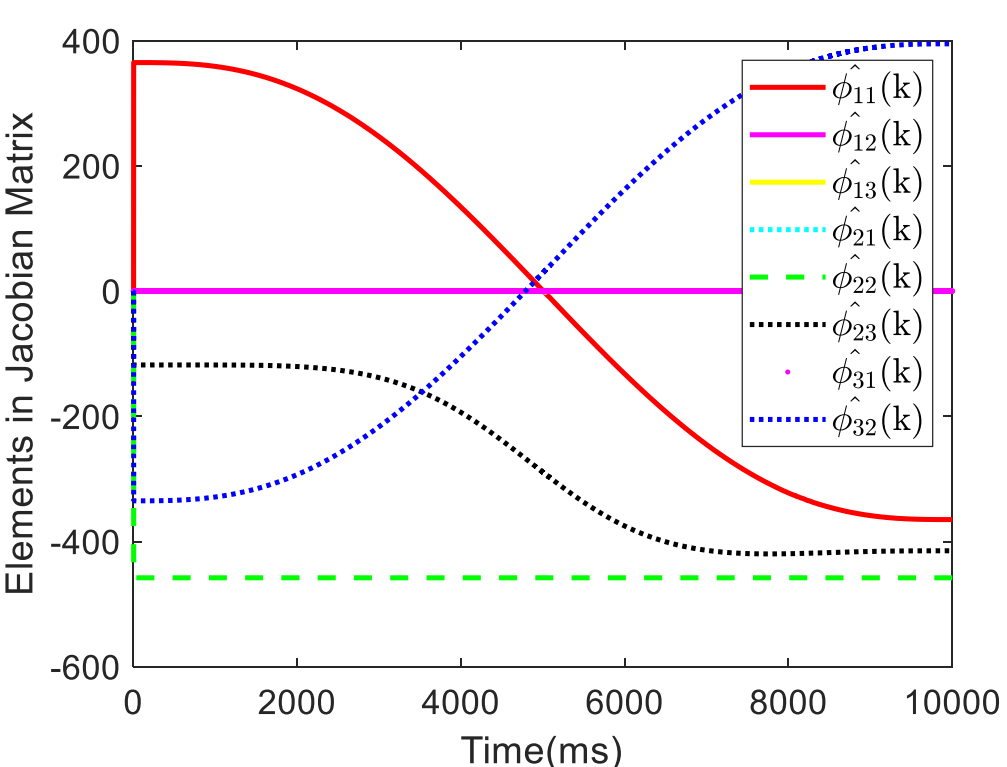

Fig. 10 Elements in $\hat{\boldsymbol{\Phi}}(k)$

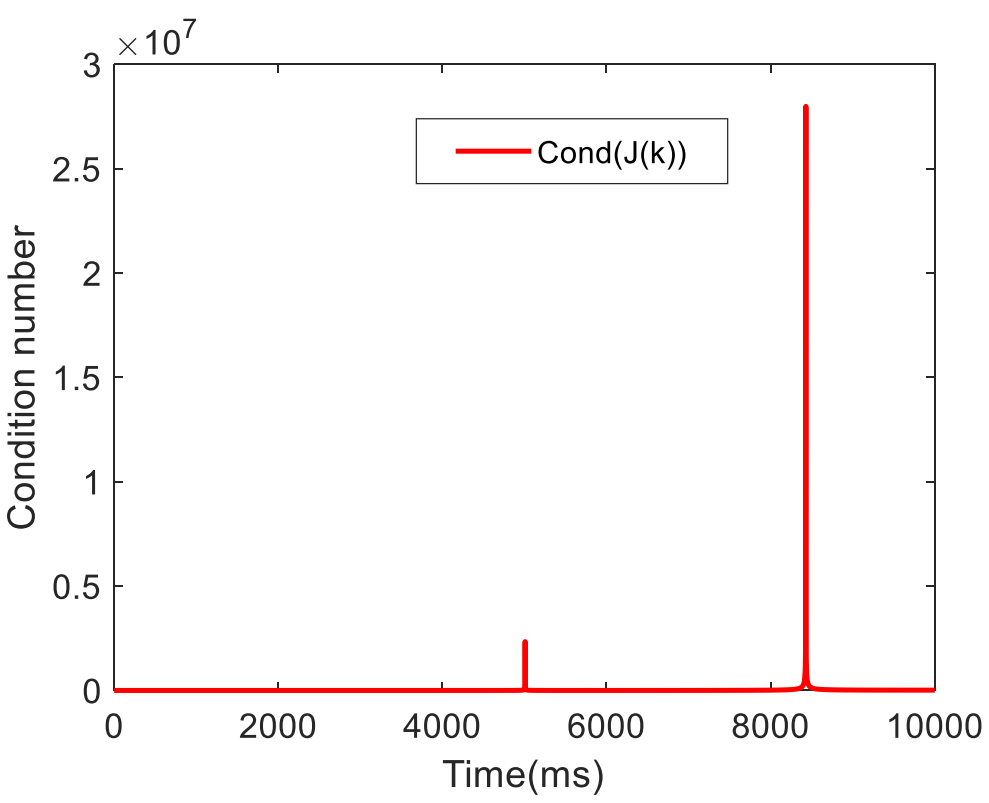

Fig. 11 Condition number in iterations

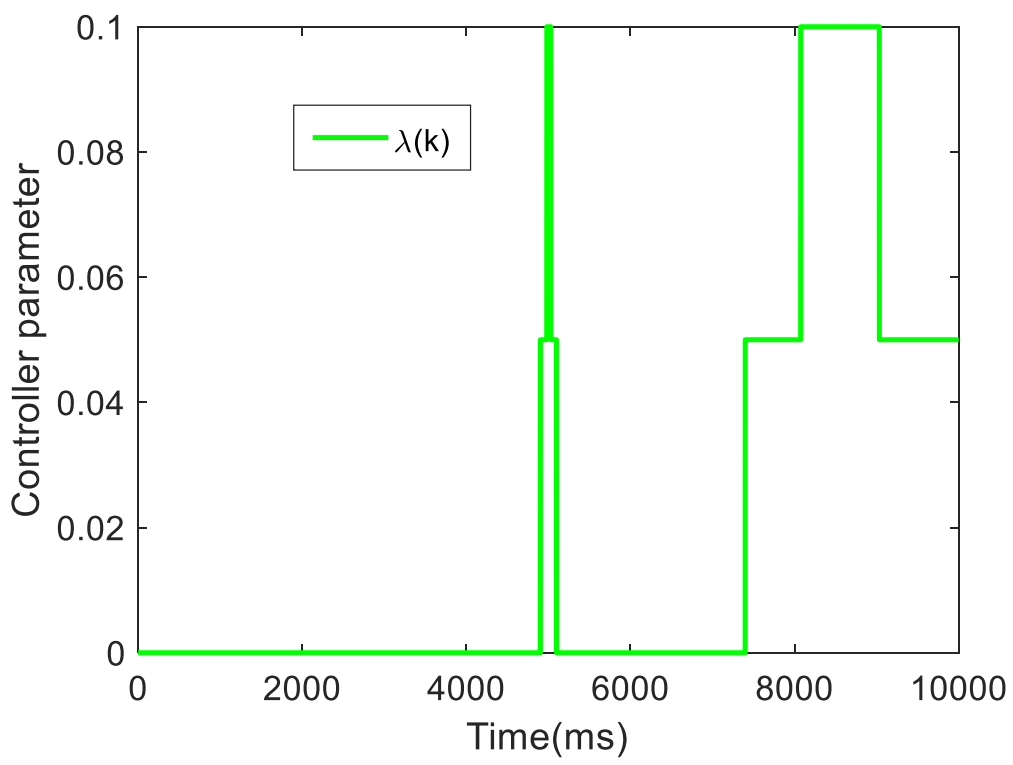

Fig. 12 Controller parameter λ

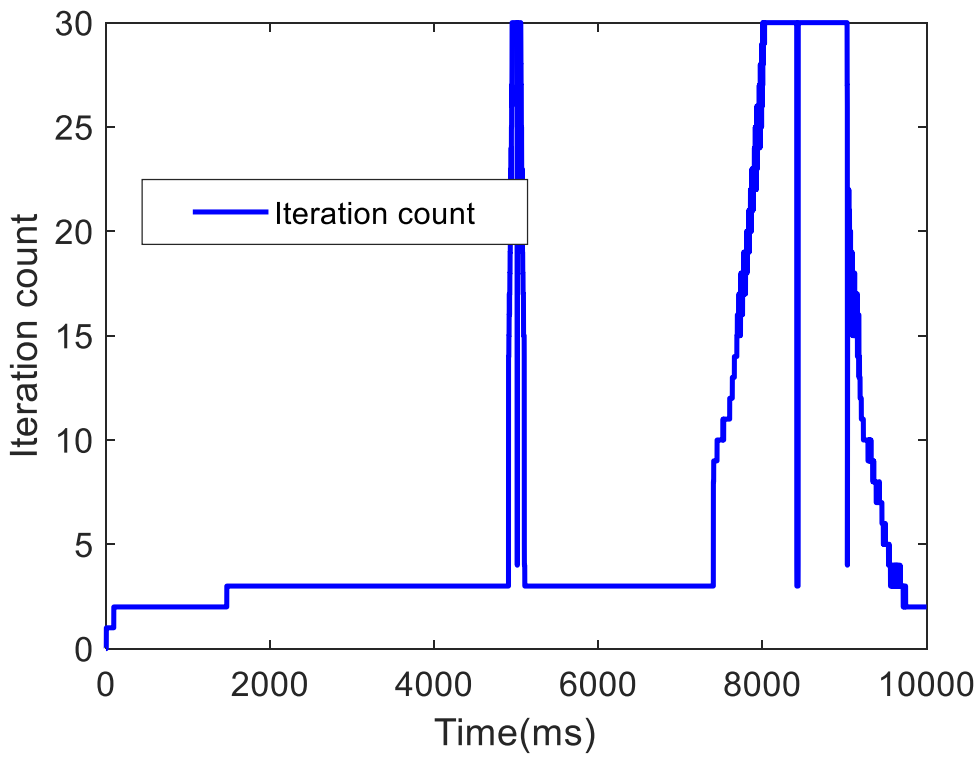

Fig. 13 Iteration count

Fig. 11 shows that the PJM is ill-conditioned at 5.009s and 8.430s. By increasing the controller parameter $\lambda$, as shown in Figure 12, the system avoids divergence but reaches the maximum iteration count of 30. Additionally, Figure 8 reveals that the controller induces the largest position errors within 0.9mm and the largest orientation errors within $4\times10^{-3}$rad when the robot operates near the singularity.

## V. CONCLUSION

In this brief, we have figured out that some MFAC methods are not studied in the right way. To this end, a kind of MFAC for a family of multivariable nonlinear systems is redesigned in this brief, and then the stability of the system and the chosen parameter $\boldsymbol{\lambda}$ are analyzed by the closed-loop function. At last, the MFAC is applied in the robotic system to demonstrate its effectiveness and, meanwhile, to exhibit an appropriate and successful application in the nonlinear system. Considering the nature of MFAC, it is necessary to rename it as the incremental one-step-ahead optimal controller.

## APPENDIX

Proof of *Theorem 1*

*Proof*: Case 1: $1\leq L_y\leq n_y$ and $1\leq L_u\leq n_u$

From (1), we have

$$
\begin{aligned}
&\Delta\boldsymbol{y}(k+1)=\\
&\boldsymbol{f}(\boldsymbol{y}(k),\cdots,\boldsymbol{y}(k-L_y+1),\boldsymbol{y}(k-L_y),\cdots,\boldsymbol{y}(k-n_y),\boldsymbol{u}(k),\\
&\quad\cdots,\boldsymbol{u}(k-L_u+1),\boldsymbol{u}(k-L_u),\cdots,\boldsymbol{u}(k-n_u))\\
&-\boldsymbol{f}(\boldsymbol{y}(k-1),\cdots,\boldsymbol{y}(k-L_y),\boldsymbol{y}(k-L_y),\cdots,\boldsymbol{y}(k-n_y),\boldsymbol{u}(k-1),\\
&\quad\cdots,\boldsymbol{u}(k-L_u),\boldsymbol{u}(k-L_u),\cdots,\boldsymbol{u}(k-n_u))\\
&+\boldsymbol{f}(\boldsymbol{y}(k-1),\cdots,\boldsymbol{y}(k-L_y),\boldsymbol{y}(k-L_y),\cdots,\boldsymbol{y}(k-n_y),\boldsymbol{u}(k-1),\\
&\quad\cdots,\boldsymbol{u}(k-L_u),\boldsymbol{u}(k-L_u),\cdots,\boldsymbol{u}(k-n_u))\\
&-\boldsymbol{f}(\boldsymbol{y}(k-1),\cdots,\boldsymbol{y}(k-L_y),\boldsymbol{y}(k-L_y-1),\cdots,\boldsymbol{y}(k-n_y-1),\\
&\boldsymbol{u}(k-1),\cdots,\boldsymbol{u}(k-L_u),\boldsymbol{u}(k-L_u-1),\cdots,\boldsymbol{u}(k-n_u-1))
\end{aligned}
\tag{49}
$$

Based on Assumption 1 and the definition of differentiability in [18], (49) becomes

$$\Delta \boldsymbol{y}(k+1)=\frac{\partial \boldsymbol{f}(\boldsymbol{\varphi}(k-1)}{\partial \boldsymbol{y}^T(k-1)}\Delta \boldsymbol{y}(k)+\cdots+\frac{\partial \boldsymbol{f}(\boldsymbol{\varphi}(k-1)}{\partial \boldsymbol{y}^T(k-L_y)}\Delta \boldsymbol{y}(k-L_y+1)$$
$$+\frac{\partial \boldsymbol{f}(\boldsymbol{\varphi}(k-1)}{\partial \boldsymbol{u}^T(k-1)}\Delta \boldsymbol{u}(k)+\cdots+\frac{\partial \boldsymbol{f}(\boldsymbol{\varphi}(k-1)}{\partial \boldsymbol{u}^T(k-L_u)}\Delta \boldsymbol{u}(k-L_u+1)$$
$$+\boldsymbol{\varepsilon}_1(k)\Delta \boldsymbol{y}(k)+\cdots+\boldsymbol{\varepsilon}_{Ly}(k)\Delta \boldsymbol{y}(k-L_y+1)$$
$$+\boldsymbol{\varepsilon}_{Ly+1}(k)\Delta \boldsymbol{u}(k)+\cdots+\boldsymbol{\varepsilon}_{Ly+Lu}(k)\Delta \boldsymbol{u}(k-L_u+1)+\boldsymbol{\psi}(k) \tag{50}$$

where

$$\boldsymbol{\psi}(k)\triangleq \boldsymbol{f}(\boldsymbol{y}(k-1),\cdots,\boldsymbol{y}(k-L_y),\boldsymbol{y}(k-L_y),\cdots,\boldsymbol{y}(k-n_y),$$
$$\boldsymbol{u}(k-1),\cdots,\boldsymbol{u}(k-L_u),\boldsymbol{u}(k-L_u),\cdots,\boldsymbol{u}(k-n_u))$$
$$-\boldsymbol{f}(\boldsymbol{y}(k-1),\cdots,\boldsymbol{y}(k-L_y),\boldsymbol{y}(k-L_y-1),\cdots,\boldsymbol{y}(k-n_y-1),$$
$$\boldsymbol{u}(k-1),\cdots,\boldsymbol{u}(k-L_u),\boldsymbol{u}(k-L_u-1),\cdots,\boldsymbol{u}(k-n_u-1)) \tag{51}$$

$$\frac{\partial \boldsymbol{f}(\boldsymbol{\varphi}(k-1)}{\partial \boldsymbol{y}^T(k-i)}=\begin{bmatrix}\frac{\partial f_1(\boldsymbol{\varphi}(k-1))}{\partial y_1(k-i)} & \frac{\partial f_1(\boldsymbol{\varphi}(k-1))}{\partial y_2(k-i)} & \cdots & \frac{\partial f_1(\boldsymbol{\varphi}(k-1))}{\partial y_{My}(k-i)}\\ \frac{\partial f_2(\boldsymbol{\varphi}(k-1))}{\partial y_1(k-i)} & \frac{\partial f_2(\boldsymbol{\varphi}(k-1))}{\partial y_2(k-i)} & \cdots & \frac{\partial f_2(\boldsymbol{\varphi}(k-1))}{\partial y_{My}(k-i)}\\ \vdots & \vdots & \vdots & \vdots\\ \frac{\partial f_{My}(\boldsymbol{\varphi}(k-1))}{\partial y_1(k-i)} & \frac{\partial f_{My}(\boldsymbol{\varphi}(k-1))}{\partial y_2(k-i)} & \cdots & \frac{\partial f_{My}(\boldsymbol{\varphi}(k-1))}{\partial y_{My}(k-i)}\end{bmatrix},$$

$$\frac{\partial \boldsymbol{f}(\boldsymbol{\varphi}(k-1))}{\partial \boldsymbol{u}^T(k-j)}=\begin{bmatrix}\frac{\partial f_1(\boldsymbol{\varphi}(k-1))}{\partial u_1(k-j)} & \frac{\partial f_1(\boldsymbol{\varphi}(k-1))}{\partial u_2(k-j)} & \cdots & \frac{\partial f_1(\boldsymbol{\varphi}(k-1))}{\partial u_{Mu}(k-j)}\\ \frac{\partial f_2(\boldsymbol{\varphi}(k-1))}{\partial u_1(k-i)} & \frac{\partial f_2(\boldsymbol{\varphi}(k-1))}{\partial u_2(k-i)} & \cdots & \frac{\partial f_2(\boldsymbol{\varphi}(k-1))}{\partial u_{Mu}(k-j)}\\ \vdots & \vdots & \vdots & \vdots\\ \frac{\partial f_{My}(\boldsymbol{\varphi}(k-1))}{\partial u_1(k-i)} & \frac{\partial f_{My}(\boldsymbol{\varphi}(k-1))}{\partial u_2(k-i)} & \cdots & \frac{\partial f_{My}(\boldsymbol{\varphi}(k-1))}{\partial u_{Mu}(k-j)}\end{bmatrix},$$

($1\le i\le L_y$, $1\le j\le L_u$) denote the partial derivative of $\boldsymbol{f}(\cdots)$ with respect to the $i$-th vector and the ($n_y$+1+$j$)-th vector, respectively. And $\boldsymbol{\varepsilon}_1(k),\cdots,\boldsymbol{\varepsilon}_{Ly+Lu}(k)$ are functions that depend only on $\Delta\boldsymbol{y}(k),\cdots,\Delta\boldsymbol{y}(k-L_y+1),\Delta\boldsymbol{u}(k),\cdots,\Delta\boldsymbol{u}(k-L_u+1)$, with $(\boldsymbol{\varepsilon}_1(k),\cdots,\boldsymbol{\varepsilon}_{Ly+Lu}(k))\to(\mathbf{0},\cdots,\mathbf{0})$ when $(\Delta\boldsymbol{y}(k),\cdots,\Delta\boldsymbol{y}(k-L_y+1),\Delta\boldsymbol{u}(k),\cdots,\Delta\boldsymbol{u}(k-L_u+1))\to(\mathbf{0},\cdots,\mathbf{0})$. This means that $(\boldsymbol{\varepsilon}_1(k),\cdots,\boldsymbol{\varepsilon}_{Ly+Lu}(k))$ will be regarded as $(\mathbf{0},\cdots,\mathbf{0})$ if the control period of the system is sufficiently small.

We consider the following equation with the vector $\boldsymbol{\eta}(k)$ for each time $k$:

$$\boldsymbol{\psi}(k)=\boldsymbol{\eta}^T(k)\Delta\boldsymbol{H}(k) \tag{52}$$

Owing to $\|\Delta\boldsymbol{H}(k)\|\neq 0$, (52) must have at least one solution $\boldsymbol{\eta}_0^T(k)$. Let

$$\boldsymbol{\phi}_L^T(k)=\boldsymbol{\eta}_0^T(k)+[\frac{\partial \boldsymbol{f}(\boldsymbol{\varphi}(k-1)}{\partial \boldsymbol{y}^T(k-1)}+\boldsymbol{\varepsilon}_1(k),\cdots,\frac{\partial \boldsymbol{f}(\boldsymbol{\varphi}(k-1)}{\partial \boldsymbol{y}^T(k-L_y)}+\boldsymbol{\varepsilon}_{Ly}(k),$$
$$\frac{\partial \boldsymbol{f}(\boldsymbol{\varphi}(k-1)}{\partial \boldsymbol{u}^T(k-1)}+\boldsymbol{\varepsilon}_{Ly+1}(k),\cdots,\frac{\partial \boldsymbol{f}(\boldsymbol{\varphi}(k-1)}{\partial \boldsymbol{u}^T(k-L_u)}+\boldsymbol{\varepsilon}_{Ly+Lu}(k)] \tag{53}$$

Then (50) can be described as

$$\Delta\boldsymbol{y}(k+1)=\boldsymbol{\phi}_L^T(k)\Delta\boldsymbol{H}(k) \tag{54}$$

Case 2: $L_y$=$n_y$+1 and $L_u$=$n_u$ +1

On the basis of *Assumption 1* and the definition of differentiability in [18], (1) becomes

$$\Delta \boldsymbol{y}(k+1)=\frac{\partial \boldsymbol{f}(\boldsymbol{\varphi}(k-1))}{\partial \boldsymbol{y}^T(k-1)}\Delta \boldsymbol{y}(k)+\cdots+\frac{\partial \boldsymbol{f}(\boldsymbol{\varphi}(k-1))}{\partial \boldsymbol{y}^T(k-n_y-1)}\Delta \boldsymbol{y}(k-n_y)$$
$$+\frac{\partial \boldsymbol{f}(\boldsymbol{\varphi}(k-1))}{\partial \boldsymbol{u}^T(k-1)}\Delta \boldsymbol{u}(k)+\cdots+\frac{\partial \boldsymbol{f}(\boldsymbol{\varphi}(k-1))}{\partial \boldsymbol{u}^T(k-n_u-1)}\Delta \boldsymbol{u}(k-n_u)$$
$$+\boldsymbol{\gamma}(k) \tag{55}$$

where

$$\boldsymbol{\gamma}(k)=\boldsymbol{\varepsilon}_1(k)\Delta\boldsymbol{y}(k)+\cdots+\boldsymbol{\varepsilon}_{Ly}(k)\Delta\boldsymbol{y}(k-n_y)$$
$$+\boldsymbol{\varepsilon}_{Ly+1}(k)\Delta\boldsymbol{u}(k)+\cdots+\boldsymbol{\varepsilon}_{Ly+Lu}(k)\Delta\boldsymbol{u}(k-n_u) \tag{56}$$

Let

$$\boldsymbol{\phi}_L^T(k)=[\frac{\partial \boldsymbol{f}(\boldsymbol{\varphi}(k-1)}{\partial \boldsymbol{y}^T(k-1)}+\boldsymbol{\varepsilon}_1(k),\cdots,\frac{\partial \boldsymbol{f}(\boldsymbol{\varphi}(k-1)}{\partial \boldsymbol{y}^T(k-n_y-1)}+\boldsymbol{\varepsilon}_{Ly}(k),$$
$$\frac{\partial \boldsymbol{f}(\boldsymbol{\varphi}(k-1)}{\partial \boldsymbol{u}^T(k-1)}+\boldsymbol{\varepsilon}_{Ly+1}(k),\cdots,\frac{\partial \boldsymbol{f}(\boldsymbol{\varphi}(k-1)}{\partial \boldsymbol{u}^T(k-n_u-1)}+\boldsymbol{\varepsilon}_{Ly+Lu}(k)] \tag{57}$$

to describe (55) as (54), with $(\boldsymbol{\varepsilon}_1(k),\cdots,\boldsymbol{\varepsilon}_{Ly+Lu}(k))\to(\mathbf{0},\cdots,\mathbf{0})$, i.e., $\boldsymbol{\phi}_L^T(k)\to[\frac{\partial \boldsymbol{f}(\boldsymbol{\varphi}(k-1))}{\partial \boldsymbol{y}^T(k-1)},\cdots,\frac{\partial \boldsymbol{f}(\boldsymbol{\varphi}(k-1))}{\partial \boldsymbol{y}^T(k-n_y-1)},\frac{\partial \boldsymbol{f}(\boldsymbol{\varphi}(k-1))}{\partial \boldsymbol{u}^T(k-1)},\cdots,\frac{\partial \boldsymbol{f}(\boldsymbol{\varphi}(k-1))}{\partial \boldsymbol{u}^T(k-n_u-1)}]$ in nonlinear systems, if $(\Delta\boldsymbol{y}(k),\cdots,\Delta\boldsymbol{y}(k-n_y),\Delta\boldsymbol{u}(k),\cdots,\Delta\boldsymbol{u}(k-n_u))\to(\mathbf{0},\cdots,\mathbf{0})$. As to linear systems, we will always have $\boldsymbol{\phi}_L^T(k)=[\frac{\partial \boldsymbol{f}(\boldsymbol{\varphi}(k-1))}{\partial \boldsymbol{y}^T(k-1)},\cdots,\frac{\partial \boldsymbol{f}(\boldsymbol{\varphi}(k-1))}{\partial \boldsymbol{y}^T(k-n_y-1)},\frac{\partial \boldsymbol{f}(\boldsymbol{\varphi}(k-1))}{\partial \boldsymbol{u}^T(k-1)},\cdots,\frac{\partial \boldsymbol{f}(\boldsymbol{\varphi}(k-1))}{\partial \boldsymbol{u}^T(k-n_u-1)}]$, no matter what $(\Delta\boldsymbol{y}(k),\cdots,\Delta\boldsymbol{y}(k-n_y),\Delta\boldsymbol{u}(k),\cdots,\Delta\boldsymbol{u}(k-n_u))$ is.

Moreover, if the function $\boldsymbol{f}(\cdots)$ has derivatives of all orders on the working points, one can obtain (58) or (59) according to the Taylor series.

$$\Delta y_t(k+1)=\frac{\partial f_t(\boldsymbol{\varphi}(k-1))}{\partial \boldsymbol{H}^T(k-1)}\Delta\boldsymbol{H}(k)+$$
$$\frac{1}{2!}\Delta\boldsymbol{H}^T(k)\frac{\partial^2 f_t(\boldsymbol{\varphi}(k-1))}{\partial\boldsymbol{H}(k-1)\partial\boldsymbol{H}^T(k-1)}\Delta\boldsymbol{H}(k)+\cdots \tag{58}$$

$$\begin{aligned}\Delta y_t(k+1) = &\Big[\sum_{p=1}^{My}\Delta y_p(k)\frac{\partial}{\partial y_p(k-1)}+\cdots+\sum_{p=1}^{My}\Delta y_p(k-n_y)\frac{\partial}{\partial y_p(k-n_y-1)}\\ &+\sum_{q=1}^{Mu}\Delta u_q(k)\frac{\partial}{\partial u_q(k-1)}+\cdots+\sum_{q=1}^{Mu}\Delta u_q(k-n_u)\frac{\partial}{\partial u_q(k-n_u-1)}\Big] f_t(\boldsymbol{\varphi}(k-1))\\ &+\cdots+\frac{1}{n!}\Big[\sum_{p=1}^{My}\Delta y_p(k)\frac{\partial}{\partial y_p(k-1)}+\cdots+\sum_{p=1}^{My}\Delta y_p(k-n_y)\frac{\partial}{\partial y_p(k-n_y-1)}\\ &+\sum_{q=1}^{Mu}\Delta u_q(k)\frac{\partial}{\partial u_q(k-1)}+\cdots+\sum_{q=1}^{Mu}\Delta u_q(k-n_u)\frac{\partial}{\partial u_q(k-n_u-1)}\Big]^n f_t(\boldsymbol{\varphi}(k-1))\\ &+\cdots\end{aligned}$$

$$(t=1,\cdots,M_y) \tag{59}$$

, and then we obtain a set of solutions (16), (17) for (56).

Case 3: $L_y>n_y+1$ and $L_u>n_u+1$

On the basis of Assumption 1 and the definition of differentiability in [18], (1) becomes

$$\begin{aligned}\Delta \boldsymbol{y}(k+1) = &\frac{\partial \boldsymbol{f}(\boldsymbol{\varphi}(k-1))}{\partial \boldsymbol{y}^T(k-1)}\Delta\boldsymbol{y}(k)+\cdots+\frac{\partial \boldsymbol{f}(\boldsymbol{\varphi}(k-1))}{\partial \boldsymbol{y}^T(k-n_y-1)}\Delta\boldsymbol{y}(k-n_y)\\ &+\frac{\partial \boldsymbol{f}(\boldsymbol{\varphi}(k-1))}{\partial \boldsymbol{u}^T(k-1)}\Delta\boldsymbol{u}(k)+\cdots+\frac{\partial \boldsymbol{f}(\boldsymbol{\varphi}(k-1))}{\partial \boldsymbol{u}^T(k-n_u-1)}\Delta\boldsymbol{u}(k-n_u)\\ &+\boldsymbol{\varepsilon}_1(k)\Delta\boldsymbol{y}(k)+\cdots+\boldsymbol{\varepsilon}_{n_y+1}(k)\Delta\boldsymbol{y}(k-n_y)\\ &+\boldsymbol{\varepsilon}_{Ly+1}(k)\Delta\boldsymbol{u}(k)+\cdots+\boldsymbol{\varepsilon}_{Ly+n_u+1}(k)\Delta\boldsymbol{u}(k-n_u)\end{aligned} \tag{60}$$

Define

$$\begin{aligned}\boldsymbol{\gamma}(k) = &\boldsymbol{\varepsilon}_1(k)\Delta\boldsymbol{y}(k)+\cdots+\boldsymbol{\varepsilon}_{n_y+1}(k)\Delta\boldsymbol{y}(k-n_y)\\ &+\boldsymbol{\varepsilon}_{Ly+1}(k)\Delta\boldsymbol{u}(k)+\cdots+\boldsymbol{\varepsilon}_{Ly+n_u+1}(k)\Delta\boldsymbol{u}(k-n_u)\end{aligned} \tag{61}$$

We consider the following equation with the vector $\boldsymbol{\eta}(k)$ for each time $k$:

$$\boldsymbol{\gamma}(k) = \boldsymbol{\eta}^T(k)\Delta\boldsymbol{H}(k) \tag{62}$$

Owing to $\|\Delta\boldsymbol{H}(k)\|\neq 0$, (62) must have at least one solution $\boldsymbol{\eta}_0^T(k)$. Let

$$\begin{aligned}\boldsymbol{\phi}_L^T(k) = \boldsymbol{\eta}_0^T(k) + \Big[&\frac{\partial \boldsymbol{f}(\boldsymbol{\varphi}(k-1))}{\partial \boldsymbol{y}^T(k-1)},\cdots,\frac{\partial \boldsymbol{f}(\boldsymbol{\varphi}(k-1))}{\partial \boldsymbol{y}^T(k-n_y-1)},\boldsymbol{0},\cdots,\boldsymbol{0},\\ &\frac{\partial \boldsymbol{f}(\boldsymbol{\varphi}(k-1))}{\partial \boldsymbol{u}^T(k-1)},\cdots,\frac{\partial \boldsymbol{f}(\boldsymbol{\varphi}(k-1))}{\partial \boldsymbol{u}^T(k-n_u-1)},\boldsymbol{0},\cdots,\boldsymbol{0}\Big]^T\end{aligned} \tag{63}$$

Then (60) can be described as (54).

Case 4: $L_y\geq n_y+1$ and $1\leq L_u<n_u+1$; $0\leq L_y<n_y+1$ and $L_u\geq n_u+1$.

The proof of Case 4 is similar to the above process, and we omit it.

We finished the proof of *Theorem 1*.

*Remark* 2: The UD in Case 2 ($L_y=n_y+1$, $L_u=n_u+1$) is as follows. (55) can be rewritten as

$$\begin{aligned}\boldsymbol{y}(k) = &\frac{\partial \boldsymbol{f}(\boldsymbol{\varphi}(k-1))}{\partial \boldsymbol{y}^T(k-1)}\boldsymbol{y}(k-1)+\cdots+\frac{\partial \boldsymbol{f}(\boldsymbol{\varphi}(k-1))}{\partial \boldsymbol{y}^T(k-n_y-1)}\Delta\boldsymbol{y}(k-n_y-1)\\ &+\frac{\partial \boldsymbol{f}(\boldsymbol{\varphi}(k-1))}{\partial \boldsymbol{u}^T(k-1)}\boldsymbol{u}(k-1)+\cdots+\frac{\partial \boldsymbol{f}(\boldsymbol{\varphi}(k-1))}{\partial \boldsymbol{u}^T(k-n_u-1)}\Delta\boldsymbol{u}(k-n_u-1)+\boldsymbol{v}(k)\end{aligned} \tag{64}$$

where

$$\begin{aligned}\boldsymbol{v}(k) = &\boldsymbol{y}(k+1)-\frac{\partial \boldsymbol{f}(\boldsymbol{\varphi}(k-1)}{\partial \boldsymbol{y}^T(k-1)}\boldsymbol{y}(k)-\cdots-\frac{\partial \boldsymbol{f}(\boldsymbol{\varphi}(k-1)}{\partial \boldsymbol{y}^T(k-n_y-1)}\boldsymbol{y}(k-n_y)\\ &-\frac{\partial \boldsymbol{f}(\boldsymbol{\varphi}(k-1)}{\partial \boldsymbol{u}^T(k-1)}\boldsymbol{u}(k)-\cdots-\frac{\partial \boldsymbol{f}(\boldsymbol{\varphi}(k-1)}{\partial \boldsymbol{u}^T(k-n_u-1)}\boldsymbol{u}(k-n_u)-\boldsymbol{\gamma}(k)\end{aligned} \tag{65}$$

is the UD around the operating point, according to the conception in [14].

## REFERENCES

[1] Hou Z S, Xiong S S. On Model Free Adaptive Control and its Stability Analysis[J]. IEEE Transactions on Automatic Control, 2019, 64(11): 4555-4569.

[2] Hou Z S, Jin S T, "A novel data-driven control approach for a class of discrete-time nonlinear systems[J]. IEEE Transaction on Control Systems Technology, 2011, 19(6):1549-1558.

[3] Hou Z S, Jin S T, Model Free Adaptive Control: Theory and Applications, CRC Press, Taylor and Francis Group, 2013

[4] Hou Z S, Jin S T. Data-Driven Model-Free Adaptive Control for a Class of MIMO Nonlinear Discrete-Time Systems[J]. IEEE Transactions on Neural Networks, 2011, 22(12): 2173-2188.

[5] Xiong S, Hou Z. Model-free adaptive control for unknown MIMO nonaffine nonlinear discrete-time systems with experimental validation[J]. IEEE Transactions on Neural Networks and Learning Systems, 2020, 33(4): 1727-1739.

[6] Zhang F L. Performance Analysis of Model-Free Adaptive Control[J]. 2020, arXiv: 2009.04248. [Online]. Available: https://arxiv.org/abs/2009.04248

[7] Zhang F L. A New Model-Free Method for MIMO Systems and Discussion on Model-Free or Model-Based[J]. 2020, arXiv:2007.02761. [Online]. Available: https://arxiv.org/abs/2007.02761

[8] Zhang F L. Discussions on Inverse Kinematics based on Levenberg-Marquardt Method and Model-Free Adaptive (Predictive) Control[J]. 2020, arXiv:2009.14507. [Online]. Available: https://arxiv.org/abs/2009.14507

[9] Zhang Y J, Jia Y, Chai T Y, et al. Data-Driven PID Controller and Its Application to Pulp Neutralization Process[J]. IEEE Transactions on Control Systems Technology, 2018, 26(3):828-841.

[10] Zhang Y J, Chai T Y, Wang H, et al. Nonlinear Decoupling Control With ANFIS-Based Unmodeled Dynamics Compensation for a Class of Complex Industrial Processes[J]. IEEE Transactions on Neural Networks & Learning Systems, 2018, 29(99):2352-2366.

[11] Zhang Y J, Chai T , Wang D H, et al. Virtual Unmodeled Dynamics Modeling for Nonlinear Multivariable Adaptive Control With Decoupling Design[J]. IEEE Transactions on Systems Man & Cybernetics Systems, 2016:1-12.

[12] Zhang Y J, Chai T Y, Wang H. A Nonlinear Control Method Based on ANFIS and Multiple Models for a Class of SISO Nonlinear Systems and Its Application[J]. IEEE Transactions on Neural Networks, 2011, 22(11):1783-1795.

[13] Zhang Y T, Chai T Y, Wang H, et al. An Improved Estimation Method for Unmodeled Dynamics Based on ANFIS and Its Application to Controller Design[J]. IEEE Transactions on Fuzzy Systems, 2013, 21(6):989-1005.

[14] Chai T Y, Yue H. Adaptive Control [M]. Beijing: Tsinghua University Press, 2015.

[15] Sugihara T. Solvability-Unconcerned Inverse Kinematics by the Levenberg–Marquardt Method[J]. IEEE Transactions on Robotics, 2011, 27(5):984-991.

[16] Lynch K M, Park F C. Modern Robotics [M]. Cambridge University Press, 2017.

[17] Craig J J. Introduction to Robotics: Mechanics and Control, Addison-Wesley, Boston, 2005.

[18] Briggs W, Cochran L, Gillett B. Calculus [M]. Pearson, 2014.